\documentclass[prd,nofootinbib,preprint,superscriptaddress]{revtex4}
\pdfoutput=1
\usepackage[utf8]{inputenc}
\usepackage{feynmp}
\usepackage{tikz-feynman}
\tikzfeynmanset{compat=1.1.0}
\usepackage{amsmath, amssymb, amsthm, graphicx, fancyhdr,epsfig, slashed, mathrsfs}
\usepackage{subfigure}
\usepackage{tikzsymbols}
\usepackage{natbib}
\usepackage{float}
\usepackage{lipsum}
\usepackage{mathtools} 
\usepackage{setspace}
\usepackage{tikz,xcolor,hyperref}
\hypersetup{
     colorlinks=true,
     citecolor    = blue,
     linkcolor= blue,
     urlcolor = blue
}

\definecolor{lime}{HTML}{A6CE39}
\DeclareRobustCommand{\orcidicon}{
	\begin{tikzpicture}
	\draw[lime, fill=lime] (0,0) 
	circle [radius=0.2] 
	node[white] {{\fontfamily{qag}\selectfont \tiny ID}};
	\draw[white, fill=white] (-0.0625,0.095) 
	circle [radius=0.007];
	\end{tikzpicture}
	\hspace{-2mm}
}

\foreach \x in {A, ..., Z}{\expandafter\xdef\csname orcid\x\endcsname{\noexpand\href{https://orcid.org/\csname orcidauthor\x\endcsname}
			{\noexpand\orcidicon}}
}

\newcommand{\be}{\begin{equation}}
\newcommand{\ee}{\end{equation}}
\newcommand{\bea}{\begin{eqnarray}}
\newcommand{\eea}{\end{eqnarray}}

\newcommand{\g}{{\rm GeV}^{-1}}

\allowdisplaybreaks

\begin{document}

\title{Axion-like particle (ALP) portal freeze-in dark matter confronting ALP search experiments}

\author{Dilip Kumar Ghosh\orcidB{}}
\email{dilipghoshjal@gmail.com}
\affiliation{School of Physical Sciences, Indian Association for the Cultivation of Science,\\  2A $\&$ 2B Raja S.C. Mullick Road, Kolkata 700032, India}

\author{Anish Ghoshal\orcidA{}}
\email{anish.ghoshal@fuw.edu.pl}
\affiliation{Institute of Theoretical Physics, Faculty of Physics, University of Warsaw, \\ ul. Pasteura 5, 02-093 Warsaw, Poland}

\author{Sk Jeesun\orcidC{}}
\email{skjeesun48@gmail.com}
\affiliation{School of Physical Sciences, Indian Association for the Cultivation of Science,\\ 2A $\&$ 2B Raja S.C. Mullick Road, Kolkata 700032, India}

\begin{abstract}
\setstretch{1.3}
The relic density of Dark Matter (DM) in the freeze-in scenario is highly dependent on the evolution history of the universe
and changes significantly in a non-standard (NS) cosmological framework prior to Big Bang Nucleosynthesis (BBN).
In this scenario, an additional  species dominates the energy budget of the universe at early times (before BBN),
resulting in a larger cosmological expansion rate at a given temperature compared to the standard radiation-dominated (RD) universe.
To investigate the production of DM in the freeze-in scenario, we consider both standard RD and NS
cosmological picture before BBN and perform a comparative analysis. We extend the Standard Model (SM) particle content with a
SM singlet DM particle $\chi $ and an axion-like particle (ALP) $a$. 
The interactions between ALP, SM particles,    
and DM are generated by higher dimensional effective operators. This setup allows the production of DM $\chi$ from SM bath
through the mediation of ALP, via ALP-portal processes.  These interactions involve non-renormalizable operators, leading to
ultraviolet (UV) freeze-in, which depends on the reheating temperature ($T_{RH}$) of the early universe. In the NS cosmological
scenario, the faster expansion rate suppresses the DM production processes, allowing for enhanced effective couplings between
the visible and dark sectors to satisfy the observed DM abundance compared to RD scenario. This improved coupling
increases the detection prospects for freeze-in DM via the ALP-portal, which is otherwise challenging to detect in RD universe
due to small couplings involved. Using an effective field theory set-up, we show that various ALP searches  such as in
FASER, DUNE, and SHiP, etc. will be able to probe significant parameter space depending on the different model parameters. 
\end{abstract}
\maketitle

\section{Introduction}
\label{sec:intro}
A variety of independent astrophysical observations have established 
the presence of an ample non-baryonic as well as electrically neutral 
dark matter (DM) constituent in our 
universe \cite{Zwicky:1933gu,Rubin:1970zza,Clowe:2006eq,Planck:2018vyg}.
Additionally, from cosmological point of view, the presence of 
DM component of the universe is expected to play a significant role in the large scale structure formation of our
universe \cite{Planck:2018vyg}. 
However, the origin and composition of the particle nature of 
DM hypothesis is a modern day intriguing puzzle for both theoretical and 
experimental particle physics world so to say. 
Since the standard model (SM) with its current particle 
contents can not accommodate DM, several well motivated beyond the SM (BSM)
scenarios suggest a suitable candidate for DM \cite{Jungman:1995df,Bertone:2004pz,Feng:2010gw}. 
One such popular 
candidate is the weakly interacting massive particle (WIMP),
which is attracts a lot of attention in the context of 
thermal dark matter models~\cite{Arcadi:2017kky, Roszkowski:2017nbc}. The
WIMP scenario demands 
dark sector-SM interaction cross-section of 
the order of electro-weak (EW) interaction strength (also known as WIMP Miracle) \cite{Lee:1977ua,Scherrer:1985zt,Srednicki:1988ce}. 
For the past several decades multiple DM 
search experiments have provided null results, thus pushing a very 
stringent bounds on the interaction strength between DM and the 
SM particles from direct detection experiments~
\cite{PandaX-II:2017hlx,PandaX:2018wtu,XENON:2020kmp,
XENON:2018voc,LUX-ZEPLIN:2018poe,DARWIN:2016hyl}, 
indirect detection~\cite{HESS:2016mib,MAGIC:2016xys} and 
through collider searches, as in for example, at the 
Large Hadron Collider (LHC)~\cite{ATLAS:2017bfj,CMS:2017zts, 
Kahlhoefer:2017dnp}.

In order to avoid these bounds on the WIMP scenario, alternative avenues 
have been explored,
where DM abundance may have been created out of equilibrium by the 
so-called {\it freeze-in} mechanism~\cite{McDonald:2001vt, 
Hall:2009bx, Bernal:2017kxu}. In this framework, DM particle is a 
{\it feebly interacting massive particle } (FIMP) having extremely 
suppressed coupling $\lesssim {\cal O}(10^{-12} - 10^{-10})$ with 
the SM particles. Thus in this scenario, with such a tiny coupling 
with the SM particles DM particles never maintain thermal 
equilibrium with SM bath in the early Universe. 
Rather the DM particles get produced non-thermally from the 
decay or via annihilation of the SM bath particles.
Thus from an initial negligible number density DM abundance increases and freezes in after sometime resulting in a fixed relic abundance.
This freeze in scenario is broadly classified into  
(a) {\it Infra-red (IR)} freeze-in relevant at lower temperature
~\cite{McDonald:2001vt, Hall:2009bx, Chu:2011be, Bernal:2017kxu, 
Duch:2017khv, Biswas:2018aib, Heeba:2018wtf, Barman:2019lvm,Ghosh:2023ocl,Belanger:2022gqc}
and (b) {\it Ultra-violet  (UV)} freeze-in which occurs at higher temperatures 
(approximately the reheating temperature of the Universe)~\cite{ 
Elahi:2014fsa, Chen:2017kvz, Bernal:2019mhf, Biswas:2019iqm, 
Barman:2020plp, Barman:2020ifq, Bernal:2020bfj, Bernal:2020qyu, 
Barman:2021tgt}. 
The main challenging aspect of this scenario to see any 
laboratory imprints (direct searches or otherwise) due to the smallness 
of the coupling strengths (therefore feebly interacting dark sector) involved in the processes. 
However, some possibilities
have been proposed to probe such scenarios.
For example, if the DM production in the early 
Universe proceeds via the decay or scattering of thermal bath particles then feeble 
couplings associated with such decays would make the dark sector 
particles very long-lived and can be looked at the LHC and beyond 
\cite{Curtin:2018mvb,Hessler:2016kwm,Belanger:2018sti,Heeba:2019jho, Mohapatra:2019ysk, Mohapatra:2020bze, Barman:2021lot, Barman:2021yaz, Nath:2021uqb, Das:2021nqj,  Barman:2022scg}. Recently Gravitational Waves from early universe has also been proposed to test such freeze-in DM scenarios and weakly coupled dark sector in general \cite{Ghoshal:2022ruy,Berbig:2023yyy}. Other detection prospects
of freeze in dark matter in direct search experiments can be seen in \cite{Elor:2021swj,Bhattiprolu:2022sdd}.

Taking a slight detour from the ongoing discussion let us mention 
another shortcomings of the SM, which involves the strong CP problem
which states that the $\theta$-parameter of QCD is highly constrained 
$(\mid \theta \mid \lesssim 10^{-10})$ from measurements of the neutron 
electric dipole moment. It has been postulated by Peccei and Quinn (PQ) 
that to resolve this puzzle, one has to introduce a hypothetical 
particle known as {\it axion}~\cite{Peccei:1977hh, Weinberg:1977ma, Wilczek:1977pj,Kim:2008hd} which is 
a (pseudo) Nambu-Goldstone boson of a spontaneously broken global chiral 
$U(1)$ symmetry (known as PQ symmetry and denoted by $U(1)_{\rm PQ}$) 
at some high scale $f_a >> \Lambda_{\rm QCD} \approx {\cal O} (100)~{\rm MeV}$. On the other hand, the QCD anomaly explicitly breaks this 
$U(1)_{\rm PQ}$ at the quantum level, leading to a tiny QCD axion mass, 
$m_a \sim  m_\pi f_\pi/f_a$, where $m_\pi$ is the pion mass and 
$f_\pi$ is the pion decay constant. In general, QCD axion mass and the 
magnitude of its couplings to the SM particles are inversely proportional 
to the axion decay constant $f_a$. The main characteristic features of 
the QCD axion, like smallness of its mass and weak derivative couplings
with the SM particles arise from its pseudo-Nambu Goldstone property. 
However, exact couplings of the QCD axion
to the SM particles are highly model dependent and there are three well known
UV complete models of QCD axion, namely PQWW~
\cite{Peccei:1977hh,Weinberg:1977ma,Wilczek:1977pj}, 
KSVZ \cite{Kim:1979if,Shifman:1979if} and DFSZ \cite{Zhitnitsky:1980tq,Dine:1981rt}, 
which have been extensively investigated in the literature. Among these, the PQWW
axion model, where $f_a \sim v_{\rm SM}$, where $v_{\rm SM} = 246$ GeV is
the SM vev, and $m_a \sim 10 $ KeV is already ruled out from 
experimental measurement of rare kaon decays, $K^+ \to \pi^+ a $ 
\cite{Peccei:2006as}. On the other hand for large $f_a >> v_{\rm SM} $ 
the axion is very light and very weakly coupled with the SM sector and
can be long lived too. These are invisible axions and can be a  
suitable candidate for dark 
matter (DM) satisfying the observed relic density of DM 
via the misalignment mechanism~(see Refs.\cite{Dine:1982ah,Preskill:1982cy, 
Abbott:1982af, Turner:1983he} for the first studies on this topic \& see Refs. \cite{Marsh:2015xka, DiLuzio:2020wdo, Co:2019jts} for review). 
Pseudo-Nambu Goldstone bosons (pNGBs) may appear in a plethora of beyond the standard model scenarios, 
typically those having spontaneously broken one or more 
global symmetries at some high scale. Such pNGBs are broadly called {\it axion like particles }(ALP) 
. Unlike QCD axions, 
ALPs are neither required to solve the strong CP problem nor their mass and couplings are related to the
QCD anomaly. Consequently, one can treat masses and couplings of ALPs as independent quantities, thus
allowing for larger parameter space that can be probed and more scope for the study of model realizations.
Like invisible QCD axion, ALPs are also valid dark matter candidates 
\cite {Arias:2012az,Jaeckel:2014qea,Ho:2018qur,Ganguly:2022imo}
or they can as well be a portal to Dark matter (DM) \cite{Gola:2021abm,Bharucha:2022lty}. 
Moreover ALPs have been extensively studied in the area of flavour physics, collider physics, 
astrophysics, cosmology (for extensive discussions see \cite{Agrawal:2021dbo, Ertas:2021mkt} and 
references therein). ALPs being a pseudo-Nambu Goldstone 
states, their mass can be much smaller than the underlying $U(1)$ global symmetry breaking scale that
suppresses their interactions with SM sector, i.e. $m_a \ll f_a $. Thus by taking a large value of $f_a$,
one can still make ALP scenario feasible to be explored in multiple current and upcoming experimental
facilities (see for example \cite{Agrawal:2021dbo}).

Based on the aforementioned discussions, one can study the phenomenology of ALPs at the electroweak scale 
in the effective field theory (EFT) framework.
In general while doing any phenomenological study one is agnostic about the details
of the underlying ultra-violet (UV) complete physics at the high scale $\Lambda \sim 4\pi f_a$~
\cite{Georgi:1986df,Brivio:2017ije}. Following the same approach, in this picture we assume that the ALP $a$ is the SM gauge 
singlet pure pseudo-scalar with only CP conserving couplings with the SM particles and its mass $m_a\lesssim 
v_{\rm SM}$. Based on this ALP EFT framework one can study the role of ALPs in 
many interesting particle physics phenomenology. For example, ALPs can explain the current discrepancy between theoretically expected and 
measured value of the anomalous magnetic moment of Muon~\cite{Marciano:2016yhf}, recently claimed apparent resonance in a particular nuclear transition of 
${}^{8}Be $ in the Atomki pair spectrometer experiment \cite{Ellwanger:2016wfe}
and more recently observed anomalous resonance in ${}^{4} He $
\cite{Krasznahorkay:2019lyl}. In addition, many collider studies have been 
carried out to look for the signatures of both stable and decaying ALPs 
\cite{Brivio:2017ije, Bauer:2017ris, Bauer:2018uxu,Liu:2022tqn,Neuhaus:2022dhu}.
Motivated by these active researches on ALP, 
in this paper we plan to explore its role as a portal to FIMP like (Majorana) fermionic 
dark matter. For this we first introduce a Majorana fermion dark matter $\chi$ 
which is SM gauge singlet in the particle spectrum. The stability of the 
DM $(\chi)$ is obtained by imposing a discrete ${\mathcal Z}_2$ symmetry 
under which $\chi$ is odd and rest of the particles are even. 
Based on the aforementioned  EFT formalism, one can express 
interaction between ALP and DM, enabling ALP to act as the portal 
connecting DM with SM particles. It should be noted that 
the phenomenology of this ALP-mediated DM~\cite{Hochberg:2018rjs} is 
very similar to that of pseudo-scalar portal 
DM \cite{Berlin:2015wwa,Beacham:2019nyx}.
\textit{In a nut-shell} 
we first explore the ALP-SM parameter space that satisfies the correct DM relic via ALP-portal 
freeze in  mechanism 
and then constrain the same parameter space using the data from various current and upcoming ALP search experiments.

In the early universe after the end of inflation, the 
dynamics of thermal DM production and its 
decoupling from the thermal bath or non-thermal DM production are 
highly dependent on the evolution history of our universe. In the standard
cosmological framework, one assumes that from the time scale of end of 
the inflation till the onset of the Big Bang Nucleosynthesis(BBN) at 
the temperature $T_{\rm BBN} \sim {\rm few~ MeV}$ the universe is mostly 
radiation dominated (RD)~\cite{Kawasaki:2000en,Ichikawa:2005vw}. It is 
interesting to note that there is no underlying guiding principle that 
can forbid one to assume the possibility of non-standard source 
that dominates the total energy budget of the universe at early epoch 
i.e. up to pre-BBN temperature. The presence of such non-standard species 
may significantly modify the dark matter dynamics in the early universe 
leading to many interesting consequences on DM phenomenology \cite {Co:2015pka,DEramo:2017gpl,DEramo:2017ecx,DEramo:2019tit}. 
It turns out that if the total energy budget of the universe is governed 
by some non-standard species, the Hubble parameter $(H)$ at any given 
temperature is always larger than the corresponding value of the Hubble 
parameter in the standard cosmology at the same temperature. 
Thus in this
non-standard cosmological picture having larger Hubble parameter, the 
universe expands faster at earlier time (higher temperature)  and eventually at some later epoch (before $T_{\rm BBN}$) 
standard radiation density $(\rho_{\rm rad})$ 
overcomes the non-standard energy density $(\rho_{\rm NS})$ of the universe.
This phase of the universe is identified by the temperature $T_{eq}$, where 
$\rho_{\rm NS}(T_{eq}) = \rho_{\rm rad}(T_{eq})$ \cite {DEramo:2017gpl}. From this equality condition, one can assume that when the temperature $T > T_{eq}$, the universe was in the non-standard cosmological era.
 Several studies have shown that WIMP or FIMP relic estimations may differ by several orders of magnitude in non standard cosmological scenario
\cite{Co:2015pka,DEramo:2017gpl,DEramo:2017ecx,DEramo:2019tit,Barman:2021ifu, Arcadi:2021doo,Biswas:2018iny,Barman:2021lot, Barman:2022njh,Ghoshal:2021ief,Ghosh:2022fws}.
It is observed that freeze-in production of DM is drastically suppressed 
in a faster expanding universe than in RD universe and hence
larger couplings (between the visible and the dark sector) are required in order to obtain observed relic density \cite{DEramo:2019tit,Calibbi:2021fld,Barman:2021lot, Barman:2022njh}.
In this paper we 
allude to this alternative cosmological pathway that 
may give rise to observed DM relic abundance for larger couplings with the 
visible sector, and thus boost the detection prospects for 
freeze-in DM. Furthermore, we explicitly show that assuming a non standard (NS) cosmological scenario prior to BBN increases the feasibility of testing the ALP portal DM framework in various ALP search experiments. It is worth noting that in the existing literature, the phenomenology of ALP or a generic pseudo-scalar mediated thermal dark matter \cite{Hochberg:2018rjs,Berlin:2015wwa,Beacham:2019nyx,Gola:2021abm} as well as non thermal dark matter \cite{Bharucha:2022lty} have been studied,  
considering only the standard radiation dominated (RD) universe. In contrast, we present a comparative analysis of ALP-mediated non-thermal dark matter assuming both RD and NS cosmological scenarios. This approach may lead to
significant insights into the experimental and phenomenological possibilities of building dark matter models via the ALP portal.

The paper is organized as follows. In Section \ref{sec:basic} we introduce
our model and discuss all higher dimensional operators that are relevant for
the DM phenomenology. In Section \ref{sec:fidm} we discuss the basic formalism 
for DM freeze-in production via ALP. In Section \ref{sec:relic} we summarize
constraints on ALP-DM parameter space from DM relic density. We also show
the effect of reheating temperature $(T_{RH})$ on the parameter model 
parameter space. Section \ref{sec:ns_cosmo} contains discussions on the 
DM freeze-in production non standard cosmology, while in 
Section \ref{sec:relic_ns} we study the consequences of non-standard cosmology
in the numerical estimation of DM relic density. 
We summarize various laboratory and astrophysical searches of axion like 
particles in Section \ref{sec:ALP_search}. In Section \ref {sec:dd} we
briefly mention the direct detection prospect of DM in this scenario. Finally,
we provide our conclusions in Section \ref{sec:concl}. Some important 
mathematical results are discussed in Appendices \ref{apxc} and 
\ref{apxcd}.


\section{Basic Set-up}
\label{sec:basic}

In this current frame work we extend the SM particle content by 
adding a SM gauge singlet Majorana fermion $\chi$ and an axion like 
particle (ALP) $a$ which is a pseudo Nambu-Goldstone boson of a spontaneously
broken global $U(1)$ symmetry at some energy scale $\Lambda $ which is much higher 
than the electroweak scale $v$. We also impose an additional 
${\mathcal Z}_2$ symmetry under which $\chi$ is odd and rest of the
particles are even such that $\chi$ is stable and play the role of dark 
matter candidate in this set up. At low energy, interaction of ALP $(a)$ with SM fields 
and dark matter $(\chi)$ is described by the following Effective Lagrangian 
up to dimension, $D=5$: 
\begin{eqnarray}
\mathcal{L}_{eff} = \mathcal{L}_{\rm ALP} + \mathcal{L}_{\rm ALP-SM}+\mathcal{L}_{\rm ALP-DM} 
\label{eq:lag_eff}
\end{eqnarray}
where, $\mathcal{L}_{\rm ALP}, \mathcal{L}_{\rm ALP-SM} $ and 
$\mathcal{L}_{\rm ALP-DM} $ denote the effective Lagrangian involving ALP only,
ALP with SM fields and ALP with dark matter respectively. 
The couplings of $a$ to the SM fields and DM arise after integrating out heavy 
fields at scale $\Lambda \sim 4\pi f_a $ producing interactions through 
dimension 5 operators
\cite{Brivio:2017ije, Georgi:1986df}:
\begin{eqnarray}
\mathcal{L}_{\rm ALP}&=& \frac{1}{2} \partial_{\mu}a \partial^{\mu}a- \frac{1}{2} M_a^2 a^2 ,
\end{eqnarray}
\begin{equation}
 \mathcal{L}_{\rm ALP-DM}= \Bar{\chi}i \partial_{\mu} \gamma^{\mu}\chi-M_{\chi} \Bar{\chi^c} \chi 
 -\frac{C_{\chi}}{f_a} \, (\Bar{\chi}\gamma^{\mu}\gamma^5 \chi) \partial_{\mu}a, 
 \label{eq:dm}
\end{equation}
where, $M_{\chi}$ denotes the mass of $\chi$ and $\frac{C_{\chi}}{f_a}$ is the effective coupling of DM and $a$.The mass of ALP is denoted by $M_a$.
The last term in the above equation is crucial 
for our analysis as it describes the interaction between DM and ALP. 

The Lagrangian containing interaction of $a$ with SM can be written as:
\begin{eqnarray}
\mathcal{L}_{\rm ALP-SM }&=& -\frac{C_G}{f_a}a\,G_{a\mu \nu}\Tilde{G}^{a\mu \nu}~~(\text{Gluon dominance})\nonumber \\
&&+ i\sum_f\frac{C_{f}}{2 f_{a}}\left(m_f + \frac{y_f}{\sqrt{2}}h  \right )\Bar{f}\gamma_5 f\,a,~~(\text{Fermion dominance})
\label{eq:sm}
\end{eqnarray}
where, $G$ is $SU(3)_C$ gauge boson and  $\frac{C_i}{f_a}$,($i=G,f$) are the respective dimensionful couplings. $f$ denotes both quark and leptons with mass $m_f$ and Yukawa coupling $y_f$. The SM Higgs is denoted by $h$.
Additional $D=5$ operators involving the ALP $(a)$ and electroweak gauge sectors are $\frac{C_{B}}{f_a}a\,B_{\mu \nu}\Tilde{B}^{\mu \nu}$ and $\frac{C_{W}}{f_a}a\,W_{a \mu \nu}\Tilde{W}^{a \mu \nu}$~(before EWSB) where, $W^a$ and $B$ are gauge fields corresponding to $SU(2)_L$ and $U(1)_Y$ gauge groups respectively and $C_j$ ($j\equiv W,B$) denote respective Wilson coefficients. After the EWSB, these
$D=5$ operators will lead to the effective coupling of ALP with photon $(\gamma)$, $Z$ and $W^\pm$ bosons\footnote{Apart from these operators, there can be additional $D=5$ operators 
involving DM \& SM fields like,  
$\mathcal{O}_{HH\chi\chi}= \frac{C_H}{\Lambda} H^{\dagger}H \Bar{\chi^c} \chi$, 
$\mathcal{O}_{d_m} = \frac{C_m}{\Lambda}  \Bar{\chi^c}\sigma_{\mu \nu}  \chi F^{\mu \nu}$ (magnetic 
dipole operator) and $\mathcal{O}_{d_e} = \frac{C_d}{\Lambda}  \Bar{\chi^c}\sigma_{\mu \nu} \gamma^5 \chi F^{\mu \nu}$ (electric dipole operator). However, both $\mathcal{O}_{d_m}$ and $\mathcal{O}_{d_e}$ vanish due to the Majorana nature of $\chi$.
The remaining operator $\mathcal{O}_{HH\chi\chi}$ being devoid of 
ALP field plays no role in connecting the phenomenology of ALP portal FIMP-DM scenario with the ALP search experiments. 
 Hence we ignore this operator for the rest of our discussions.}. 
For simplicity and initiatory study of freeze-in dark matter via ALP-portal we focus explicitly on fermion dominance and gluon-dominance cases and 
 neglect the presence of other terms.
The effective interactions considered in our analysis are well grounded from the UV complete models (see for example ref.\cite{Co:2019jts} (for ALP-vector boson interactions) and ref.\cite{Bharucha:2022lty} (for ALP-fermion interactions))\footnote{For UV completion also see ref.\cite{Kim:1979if,Shifman:1979if} (KSVZ model) and ref.\cite{Zhitnitsky:1980tq,Dine:1981rt}(DFSZ model).}.
For ALP-DM effective interaction see appendix A in ref.\cite{Bharucha:2022lty}.  
By adding a scalar singlet, $S$, to the SM and assigning a specific PQ charge to only $\chi$ and $S$, we can easily avoid direct DM-SM couplings at dimension 5 \cite{Bharucha:2022lty}.

The interactions mentioned in eq.\eqref{eq:sm} along with eq.\eqref{eq:dm} provide a viable connection between SM and DM via ALP portal,
and lead to the following interactions:
\begin{itemize}
\item{Freeze out from SM bath mediated by $a$ e.g. $gg\leftrightarrow  \chi \chi $, $ff \leftrightarrow \chi \chi$, $ffh \leftrightarrow \chi \chi $, $V V \leftrightarrow \chi \chi$ (where, $V\equiv W^{\pm},Z$) as well as elastic scattering like   $f \chi\leftrightarrow f \chi$,$g \chi\leftrightarrow g \chi$, $V \chi\leftrightarrow V \chi$.}
\item{Freeze in from SM bath mediated by $a$ e.g. $ff \to \chi \chi$, $gg \to \chi \chi$, $V V\to \chi \chi$ .  }
\item{Freeze out from ALPs when ALPs may or may not be in thermal bath e.g. $aa\leftrightarrow \chi \chi$.}
\item{Freeze in from ALPs when ALPs may or may not be in thermal bath e.g. $aa\rightarrow \chi \chi$, $a\rightarrow \chi \chi$.}
\end{itemize}
Detailed description of aforementioned scenarios are discussed in ref.\cite{Hambye:2019dwd, Bharucha:2022lty}.
Since ALP-SM and ALP-DM couplings are naturally suppressed it motivates us to study DM freeze-in production from the SM bath with ALP as the mediator.
We show the parameter space where both ALP and $\chi$ were not in thermal bath due to their feeble couplings. 
From the above mentioned Lagrangian in eq.\eqref{eq:sm} $\&$ eq.\eqref{eq:dm} it is understood that DM can be produced from processes like: $f \Bar{f}/gg\rightarrow  \chi \chi $, and also from the fusion of other vector bosons.
However, the same Lagrangian can lead to different two step production mechanisms 
like $f \Bar{f}/gg\rightarrow a a$ followed by $a a \rightarrow \chi \chi $.
This mechanism of DM production is often named as ``{\it sequential freeze-in}" in 
literature \cite{Hambye:2019dwd}.
However, we restrict our parameter space where DM relic contributions from the processes like  $f \Bar{f}/gg\rightarrow a a$ with $aa \rightarrow \chi \chi $ are negligible compared to direct freeze-in (mediated via ALP) \footnote{In direct freeze in from gluon fusion DM relic is proportional to $(C_G/f_a)^2.(C_\chi/f_a)^2$, whereas in the case of ``{\it sequential freeze-in}", DM relic scales as $(C_\chi/f_a)^4$. Thus ``{\it sequential freeze-in}" dilutes the connection of DM production and late time observables (which  rely on ALP-SM couplings  not on the ALP-DM coupling) \cite{Bharucha:2022lty,Hambye:2018qjv}. Hence to avoid {\it sequential freeze-in} process in this framework, we assume $C_G/f_a,~C_f/f_a\gg C_\chi/f_a$ throughout the analysis.}. 
The {\it sequential freeze-in} is less effective from the perspective of laboratory searches (for ALP as the mediator) of our scenario 
which will be addressed in the later part (sec.\ref{sec:relic}) of this paper.

\begin{figure}
\subfigure[\label{fd1}]{
    \includegraphics[scale=0.5]{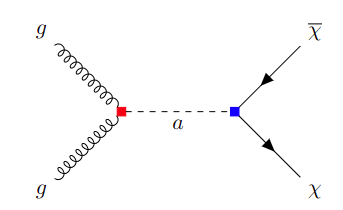}}
\subfigure[\label{fd2}]{
    \includegraphics[scale=0.5]{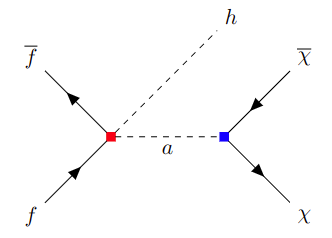}}    
\caption{\it Feynman diagrams involved in ALP-portal freeze in of DM of $\chi$ assuming (a) gluon dominance and (b) fermion dominance.}

\label{fig:1}
\end{figure}

In principle all five operators may be present at the same time in the DM production process. Simultaneous presence of several such operators not only complicate the numerical analysis but also hide the crucial underlying dynamics of DM production mechanism we allude to explore. Hence
for simplicity we will consider one effective operator at a time describing the ALP-SM interaction to understand the dependencies of the involved parameters in a more concrete manner. We will focus in details  on the following two cases: 
\begin{enumerate}
  \item {\bf Gluon Dominance:} in this case we consider DM production only 
from gluon fusions and assume $C_B/f_a=C_W/f_a=C_f/f_a=0, C_G/f_a \neq 0 $. Here the freeze in production channel will be $gg\to \chi \chi$ only.
  \item {\bf Fermion Dominance:} in this case we consider DM production from 
fermions only assuming $C_B/f_a=C_W/f_a=C_G/f_a=0, C_f/f_a \neq 0 $. 
Here the production channels will be  $f \Bar{f}\rightarrow  \chi \chi$  and 
$f \Bar{f}\rightarrow  \chi \chi h$. Before electroweak 
symmetry breaking (EWSB)($T_{\rm EW}$) only $f \Bar{f}\rightarrow \chi \chi h$ process is available
as understood from eq.\eqref{eq:sm}.
Another production channel will be $f \Bar{f}h\rightarrow \chi \chi $ which is more phase space suppressed compared to 
$f \Bar{f}\rightarrow \chi \chi h$.
\end{enumerate} 
The Feynman diagrams corresponding to above two channels are shown in fig.\ref{fig:1}.
DM production from other vector boson fusions are perfectly possible.
However, in this work we study only the aforementioned  two possibilities and confine our analysis in more minimal set up.
Ref. \cite{Bharucha:2022lty} has also studied DM freeze in from SM bath via ALP mediated processes and considered only fermion dominance scenario. In this work we include gluon dominance scenario also and discuss the prospects of laboratory searches in the non-standard cosmological frame-work which is absent in the existing literature.
\section{Freeze-in Dark Matter}
\label{sec:fidm}

From our discussions in the previous section, it is evident that 
the freeze-in production of DM and DM relic density in the acceptable range can be induced by various 
dimension 5 operators. 
This mechanism of freeze in by higher dimensional operators is popularly known as 
ultra violet  (UV) freeze-in \cite{Elahi:2014fsa}. 
In this case, DM relic density 
not only depends on various model parameters, but also 
directly proportional to the reheating temperature ($T_{RH}$) of the visible 
sector \cite{Elahi:2014fsa}. 
$T_{RH}$ is defined as the temperature at the end of reheating process, when the energy density stored in inflaton gets transferred to the radiation bath. 
In the case of perturbative instantaneous reheating $T_{RH}$ is often approximated as the equality temperature between inflaton decay width and the Hubble parameter $H$ \cite{Kolb:1990vq,BazrafshanMoghaddam:2017xrv}.
Although, $T_{RH}$ is often assumed to be the maximum temperature obtained by thermal bath after the end of inflation \cite{Elahi:2014fsa}, it is crucial to
note that maximum temperature of the universe in principle can be larger than $T_{RH}$ as discussed in  \cite{Chung:1998rq}. 
The lower bound on $T_{RH}$ from BBN is a $\sim$ few MeV \cite{Hannestad:2004px}. 
In our EFT framework with a cut off scale $\Lambda \sim 4\pi f_a$, we assume $T_{RH}< \Lambda $ as the maximum temperature and cut-off for the 
evaluation of Boltzmann equation and also treat $T_{RH}$ as free parameter obeying the bounds discussed above
\cite{Elahi:2014fsa, Chen:2017kvz, Bernal:2019mhf,Barman:2020ifq,Barman:2020jrf}. 
The detailed analysis of $T_{RH}$ in context of freeze-in DM may depend on the choice of specific inflationary models \cite{Calibbi:2021fld,Barman:2022tzk,Ahmed:2022tfm,Barman:2023ktz} which is beyond the scope of this work. 
For the two dominant higher dimensional operators discussed earlier, the DM relic density depends on these model parameters: 
\begin{equation}
C_{\chi}/f_a,\,M_a,\,M_{\chi},\,~T_{RH},~C_G/f_a,~C_f/f_a .   
\end{equation}
With these set of parameters in hand, we now discuss the freeze-in 
production of $\chi$ for these two cases one at a time: 
\begin{itemize}
\item Gluon Dominance process 
\item Fermion Dominance process
\end{itemize}
In the first case DM ($\chi$) is produced non-thermally through 
ALP mediated gluon fusion process. To find the number density of $\chi$ 
produced in this process we solve the following Boltzmann equation,
\begin{equation}
  \dot{n}_{\chi} + 3 H n_{\chi} =c[\mathcal{F}_g], 
  \label{eq:beq}
\end{equation}
where, $c[\mathcal{F}_g]$ is the collision term corresponding to this process and 
is given by
\begin{equation}
   c[\mathcal{F}_g] = \int d\Pi_g d\Pi_g d\Pi_{\chi} d\Pi_{\chi} \delta^4(p_{1}+ 
p_{2} -p_{3}-p_{4})
    |M|^2_{\rm g (p_1) g (p_2)\to \chi (p_3) \chi (p_4)} \mathcal{F}_g(p_1) \mathcal{F}_g(p_2) ,
\end{equation}
where, $\Pi_i~(i=g,\chi)$ denote the Lorentz invariant phase space,
with initial and final state 4 momenta $p_1,p_2$ and $p_3,p_4$ respectively.
The dynamics of the process is clubbed in the matrix element 
$|M|^2_{\rm gg-\chi \chi}$ and $\mathcal{F}_g$ stands for the gluon 
distribution function.
As gluons are relativistic particles one should use the relativistic treatment and 
take $\mathcal{F}_g$ as the Bose-Einstein (BE) distribution in principle.
However,it turns out that at high temperature the thermal averaged cross-section 
differs by only factor of about 3 if we use Maxwell-Boltzmann (MB) statistics instead of BE and leads to same number at low temperature as shown in Appendix \ref{apxc}.
Since both BE and MB distribution give more or less same number density, we can use MB distribution for gluon 
to find the thermal averaged cross-section. 
Converting the number density ($n_{\chi}$) to dimensionless variable 
$Y_{\chi}$ (where ,$Y_{\chi}=n_{\chi}/s$ and $s$ is the entropy density), popularly knwon as the DM yield we get the following equation:

\begin{equation}
\dfrac{dY_{\chi}}{dT}= - \dfrac{1}{s H T}\left<\sigma v\right>_{gg\rightarrow \chi  \chi} (Y_g^{eq})^2, 
\label{eq:dm_1}
\end{equation}
 where, $Y_g^{eq}$ is the equilibrium abundance of gluon and $\left<\sigma v\right>_{gg\rightarrow \chi  \chi}$ is the 
 thermally averaged cross section $\sigma_{g g \to \chi \chi}$ given in eq.\eqref{eq:sig1}. $s$ is the co-moving entropy density and $H$ is the Hubble parameter that decides expansion rate of the universe. 
 Details of yield calculation is given in Appendix \ref{apxc}. However using simple dimensional arguments one can see dark matter yield,
Y$_{\chi}$ has the following functional dependence on various model parameters \cite{Elahi:2014fsa}:
\begin{equation}
    Y_{\chi} \sim \left(\dfrac{C_{G}}{f_a}\right)^2 \left(\dfrac{C_{\chi}}{f_a}\right)^2 ~M_{\chi}^2~ M_{pl}~   T_{RH}
    \label{eq:ydm_g}
\end{equation} 
 
\begin{figure}
    \centering
     \subfigure[\label{fig:dm_a}]{
    \includegraphics[scale=0.45]{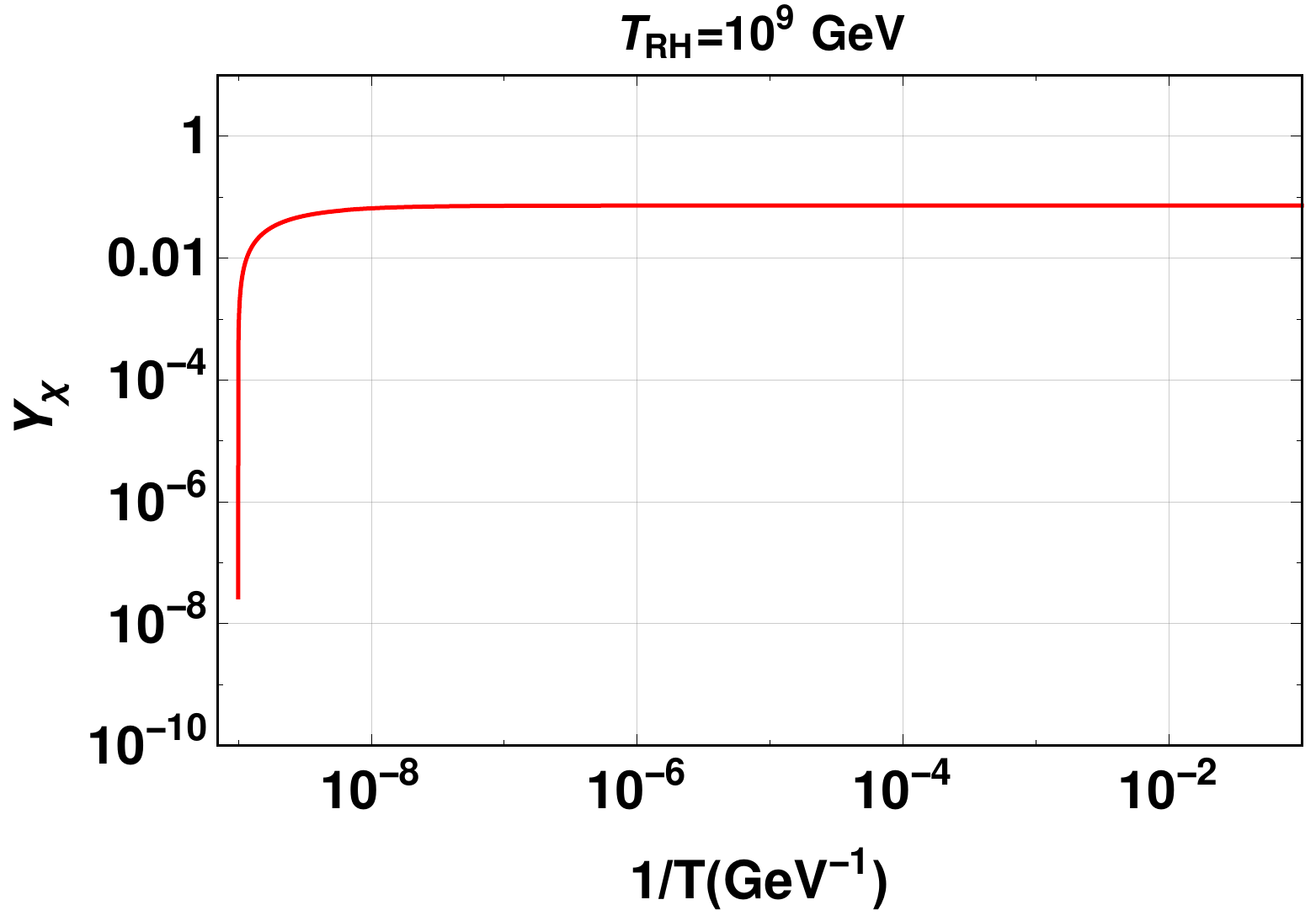}}
    \subfigure[\label{fig:dm_b}]{
    \includegraphics[scale=0.45]{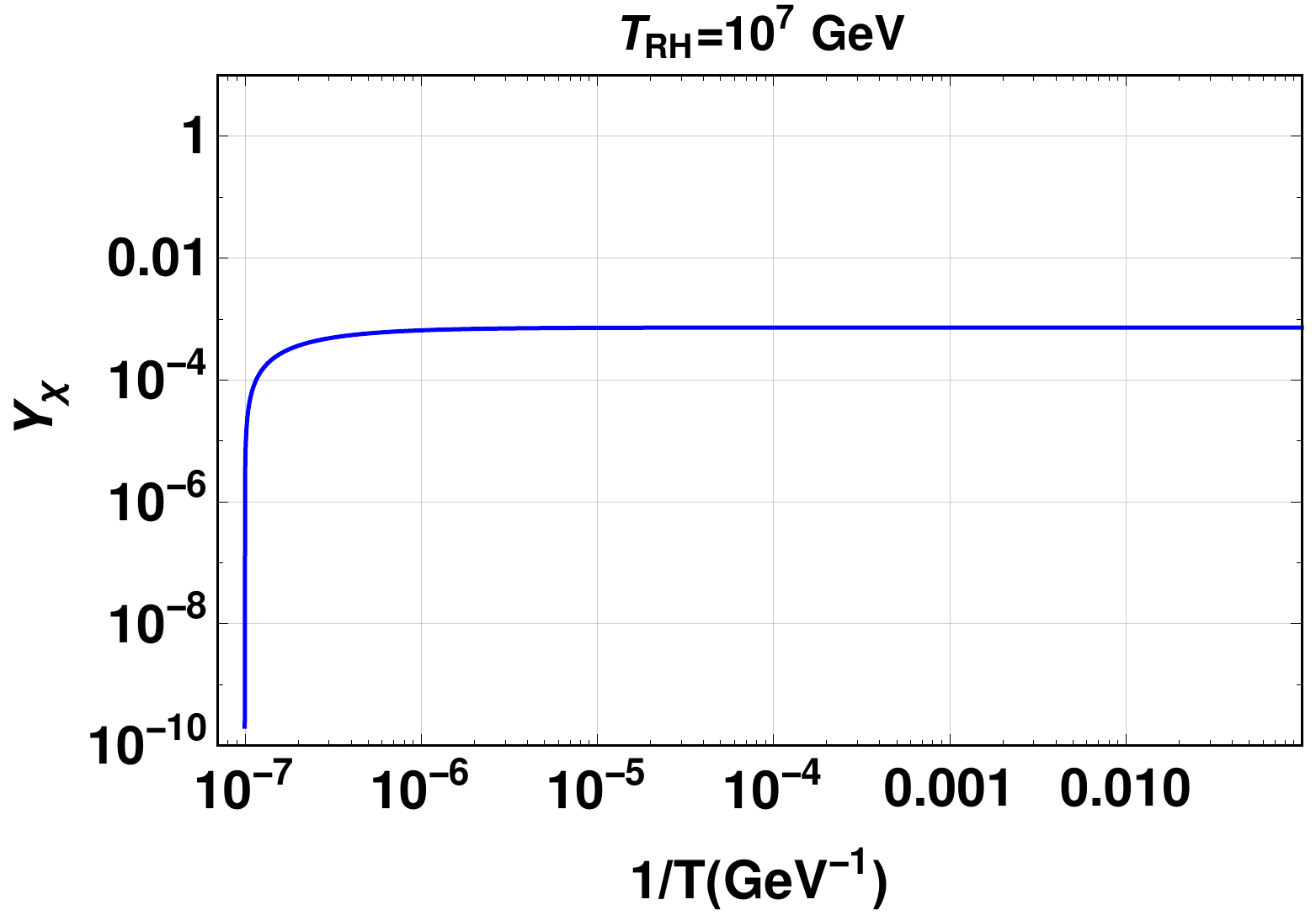}}
    \caption{\it Variation of abundance $Y_{\chi}$ with $1/T$ ({\rm GeV}$^{-1}$) with $C_{\chi}/f_a=10^{-12}~\g,~C_G/f_a=10^{-10}~\g$, $M_{\chi}=10^3$ {\rm GeV}, $M_a=0.1$ {\rm GeV} considering $T_{RH}=10^9$ {\rm GeV} in (a) and $T_{RH}=10^7$ {\rm GeV} in (b).}
    \label{fig:dm}
\end{figure}
In Fig.\ref{fig:dm}
 we display the variation of the 
co-moving abundance $Y_{\chi}$ with $1/T$
for a representative value of $C_{\chi}/f_a=10^{-12}~ \g,~C_G/f_a=10^{-10}~\g$, $M_a=0.1$ GeV and $M_{\chi}=10^3$ GeV .
We consider $T_{RH}=10^9$ and $ 10^7$ GeV in Fig.\ref{fig:dm_a} and Fig.\ref{fig:dm_b} respectively.
Comparing  Fig.\ref{fig:dm_a} and Fig.\ref{fig:dm_b} we note that with an increase in $T_{RH}$ the freeze-in takes place at earlier time with higher abundance. 
This behavior can be easily understood from the dependence of $Y_\chi$ on $T_{RH}$ in eq.\eqref{eq:ydm_g}.

Similar kind of analysis is performed with fermion dominance for 
which the collision term $C[\mathcal{F}]$ in eq.\eqref{eq:beq} involves 
processes like $f \Bar{f}\rightarrow  \chi \chi$  and 
$f \Bar{f}\rightarrow  \chi \chi h$. As realized from eq.\eqref{eq:sm}, 
that before the electroweak symmetry breaking (EWSB)($T_{\rm EW}$) 
only $f \Bar{f}\rightarrow \chi \chi h$ process is available, and
for $T\gg T_{\rm EW}$ the abundance, $Y_{\chi}$ has the following functional form \cite{Elahi:2014fsa,Bharucha:2022lty}: 
\begin{equation}
 Y_{\chi} \sim y_f^2 N_c \left(\dfrac{ C_{f}}{f_a} \right)^2 \left(\dfrac{C_{\chi}}{ f_a}\right)^2 M_{\chi}^2 ~M_{pl}  ~T_{RH},
 \label{eq:ydm_f}
\end{equation} 
where, $N_c$ is the color factor associated with quarks. The detailed numerical formulation is shown in Appendix eq.\eqref{eq:cff}. Comparing eq.\eqref{eq:ydm_g} and \eqref{eq:ydm_f} one notices that for fermion dominance $Y_{\chi}$ has same dependence on $C_f$  as on $C_g$ in gluon dominance apart from the Yukawa coupling and the color factor. 
We also have considered other production channels like $f \Bar{f}h\rightarrow \chi \chi $, 
$f h\rightarrow f \chi \chi $ and $Y_{\chi}$ takes similar form as 
in eq.\eqref{eq:ydm_f}.
 Nevertheless for fermion dominance also, DM is produced from the UV freeze-in and will have the same functional
dependence on  $T_{RH}$ as in the case of gluon dominance (shown in Fig.\ref{fig:dm}). For $T_{RH}< T_{\rm EW}$, 
$f \Bar{f}\rightarrow  \chi \chi$ process also becomes relevant. However we find that for low $T_{RH}$ the DM remains under abundant (as $Y_\chi\propto T_{RH}$) for most of the parameter space.
This motivates us to consider $T_{RH}> T_{\rm EW}$ throughout our analysis.

Since we are interested in freeze-in production of DM, the criterion of non-thermal production of DM $\chi$ gives us an upper bound on the  interaction rates.
Hence, to forbid the thermalization of $\chi$ one must satisfy the following conditions,
\begin{eqnarray}
&&\Gamma_{gg \to \chi \chi}(T_{RH})< H(T_{RH})~\text{(gluon dominance)},
\label{eq:therm}\\
&&\Gamma_{f \Bar{f} \to \chi \chi h}(T_{RH})< H(T_{RH})~\text{(fermion dominance)}.
\label{eq:thermf}
\end{eqnarray}
where, $H$ is the Hubble expansion rate ($H\sim 1.66\, T^2/ M_{pl}$) and 
$\Gamma_{gg \to \chi \chi}$ is the interaction rate of the process $gg \to \chi \chi$ given by $n_g^{eq} \left <\sigma_{gg \to \chi \chi} v\right>$ where $n_g^{eq}$ is the equilibrium number density of gluons. Similarly $\Gamma_{f \Bar{f} \to \chi \chi h}$ is the interaction rate for $f \Bar{f} \to \chi \chi h$ given by $n_f^{eq} \left <\sigma_{f \Bar{f} \to \chi \chi h} v\right>$ where $n_f^{eq}$ is the equilibrium number density of fermions.
  In this UV freeze in scenario, since most of the DM production happens at a temperature $T=T_{RH}$, we check the DM thermalization condition at $T = T_{RH}$. 
  As the interaction rate increases with increasing temperature, the annihilation process might get thermalized at high $T_{RH}$. 
  This leads to an upper bound on $T_{RH}$ beyond which $\chi$ gets thermalised and the upper limit $(T_{RH})_{\rm max}$ depends on the couplings $(C_g/f_a),~(C_f/f_a),\, (C_{\chi}/f_a) $ and $M_{\chi}$. 
  From eq.\eqref{eq:therm} and \eqref{eq:thermf} one can approximate the upper limit on $T_{RH}$ as, 
\begin{equation}
T_{RH}\lesssim \left(\frac{C_g}{f_a} \right)^{-2} \left(\frac{C_{\chi}}{f_a} \right)^{-2} M_{\chi}^{-2} M_{pl}^{-1}
~\text{(gluon dominance)},
\label{eq:therm1}    
\end{equation}
\begin{equation}
T_{RH}\lesssim \left(\frac{y_f C_f}{f_a} \right)^{-2} \left(\frac{C_{\chi}}{f_a} \right)^{-2} M_{\chi}^{-2} M_{pl}^{-1}
~\text{(fermion dominance)}.
\label{eq:therm2}    
\end{equation}
The detailed calculations are shown in appendix \ref{apxc}. 
The lower bound on $T_{RH}>$ few MeV comes from Big Bang Neucleosynthesis (BBN) \cite{Ichikawa:2005vw,Kawasaki:2000en}.

\medskip
\section{Relic Density}
\label{sec:relic}
In the previous section we have discussed that the DM ($\chi$) abundance is generated due to the non-thermal production from SM bath via ALP portal. The abundance of $\chi$ is decided by  the annihilation cross section of gluons (fermions) or in other words, by SM-ALP coupling ($C_G/f_a,~C_f/f_a$) as well as ALP-dark sector coupling ($C_{\chi}/f_a$) as understood from our earlier discussions.  
We estimate the relic density of $\chi$ after solving for $Y_{\chi}$ from eq.\eqref{eq:beq} and is given by,
\be
\Omega h^2= 2.755 \times 10^8 \, Y_{\chi} \left(\frac{M_{\chi}}{\rm GeV} \right)
\ee
 
In this UV-freeze-in process besides the couplings and $M_{\chi}$, the relic density also depends on  $T_{RH}$.
We are interested in ALP mass $M_a\sim\mathcal{O}$(MeV- 10 GeV) as this is the ball-park region that is suitable to long-lived ALP-portal searches as we will show in subsequent section. 
If the dark matter mass $M_\chi < M_a/2$, the relic density of DM receives an additional contribution from the ALP decay process $a \to \chi \chi$. However, for the sake of simplicity, 
we restrict our model parameters such that $M_\chi > M_a/2$ to forbid the aforementioned
ALP decay and annihilation of SM bath particles via the ALP portal provides the major contribution to
the DM density. One should note that there can be another contribution to the 
DM abundance from the scattering of thermalized ALPs via freeze in process \cite{Bharucha:2022lty}.
In this case, once the ALPs become thermalized, the DM relic only depends upon $C_\chi /f_a$, 
thus decoupling the DM relic from any experimental probes of ALP.
However, in order to highlight the tight connection between the ALP-portal DM 
scenario and the experimental search of ALPs at present and future laboratory 
experiments, we restrict our model parameter space such that 
$aa \leftrightarrow SM$ does not thermalize and also consider 
$ C_i/f_a>C_\chi/f_a ~(i \equiv G,f)$ to reduce such contribution to DM abundance \cite{Hambye:2019dwd}.
The freeze-in production in our case takes place mostly at $T_{RH}\sim$ a few TeV which is much higher than the mediator mass. 
In this limit the thermally averaged cross-sections, $\left<\sigma v\right>_{gg\rightarrow \chi  \chi}$, and $\left<\sigma v\right>_{\Bar{f} f\rightarrow \chi  \chi h}$  are almost independent of $M_a$. 
Thus our relic density analysis is almost insensitive  of ALP mass
\footnote{For the dependence of light mediator mass on relic density see Ref.\cite{Mohapatra:2019ysk,Barman:2021yaz}}. 
Before diving into relic density analysis and laboratory searches we want to shed some light on the allowed parameter space from the conditions of thermalization in early universe.


\begin{figure}
    \subfigure[\label{fig:th1}]{
    \includegraphics[scale=0.35]{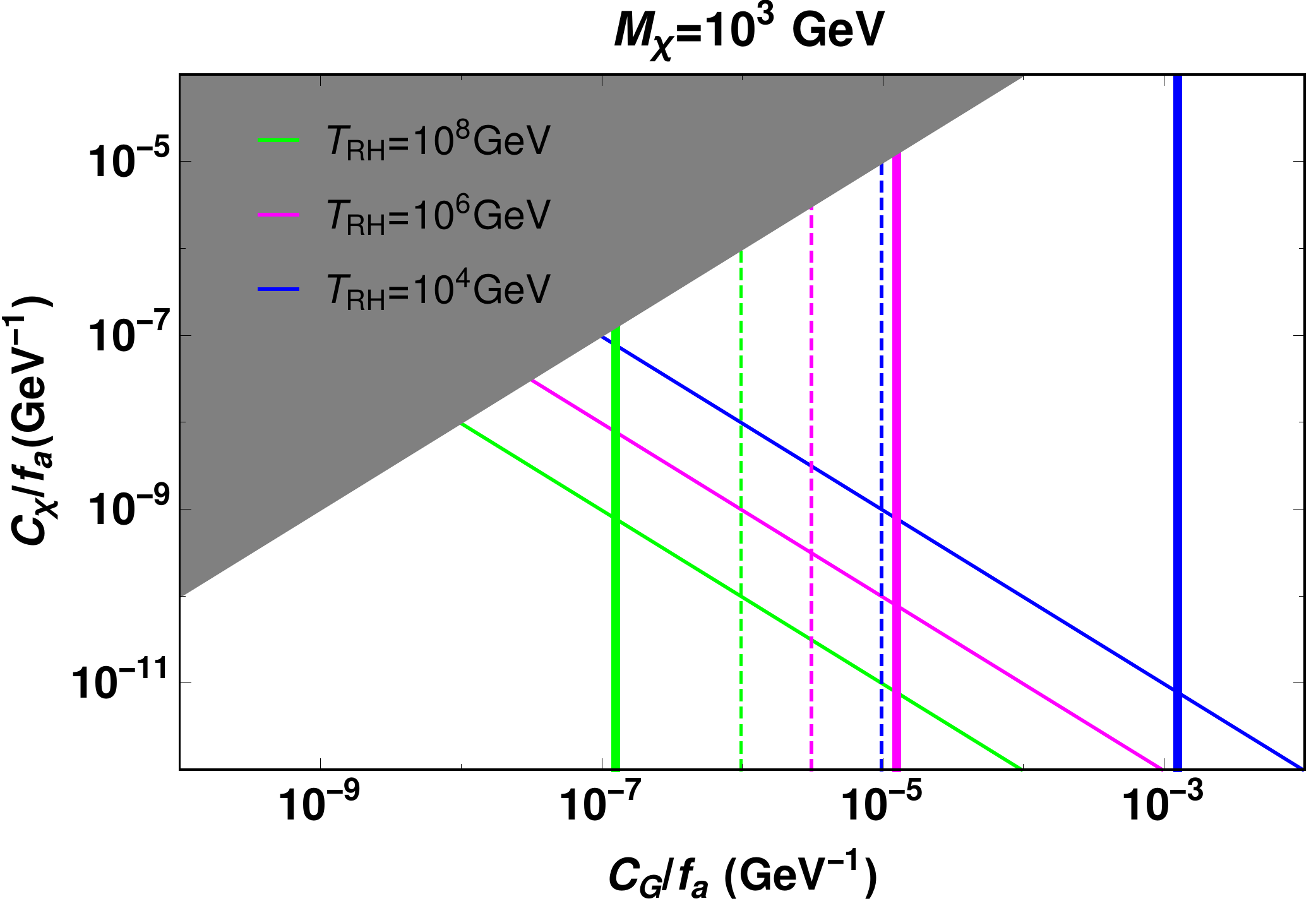}}
    \subfigure[\label{fig:th2}]{
    \includegraphics[scale=0.35]{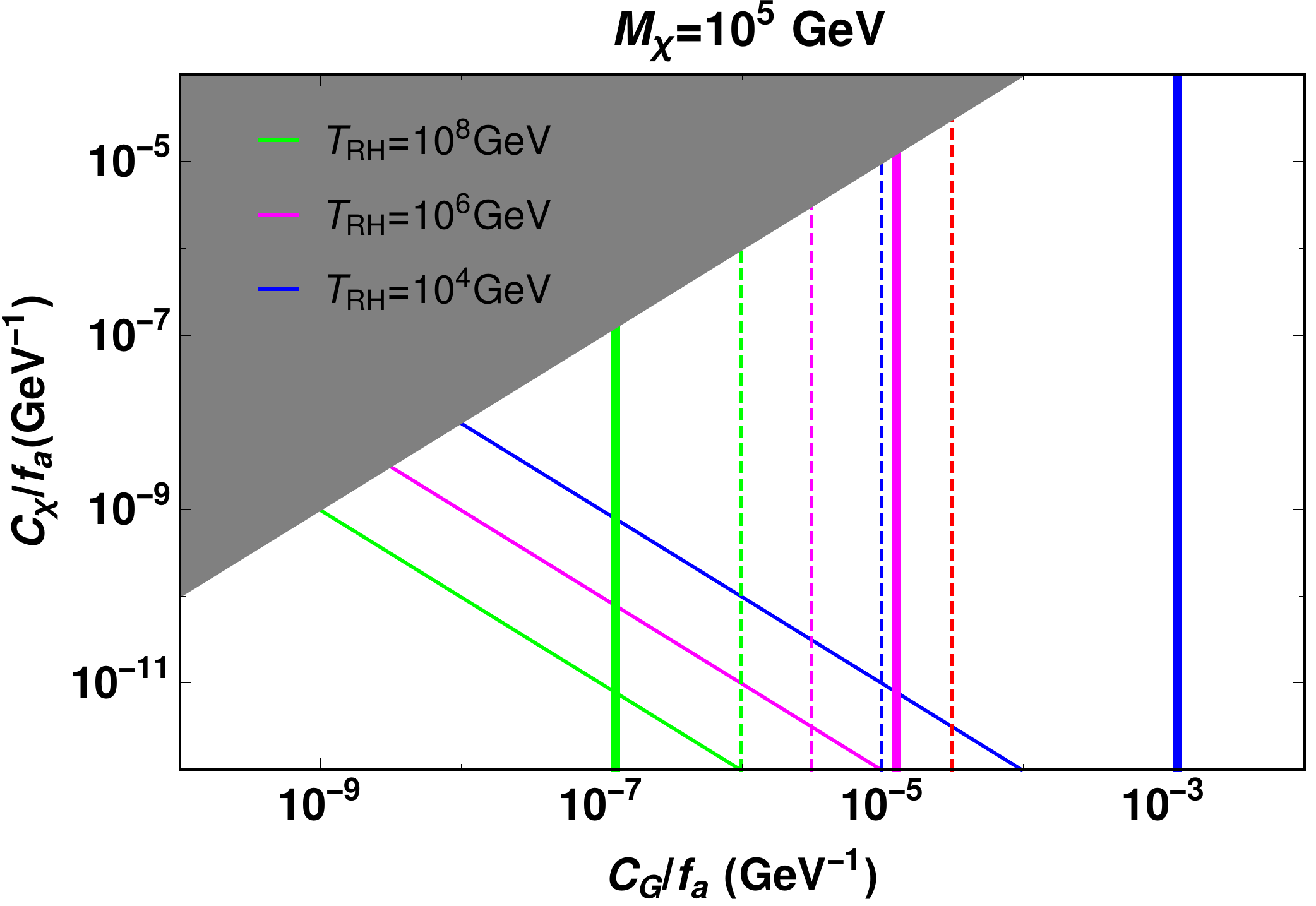}}
    \subfigure[\label{fig:f_th1}]{
    \includegraphics[scale=0.35]{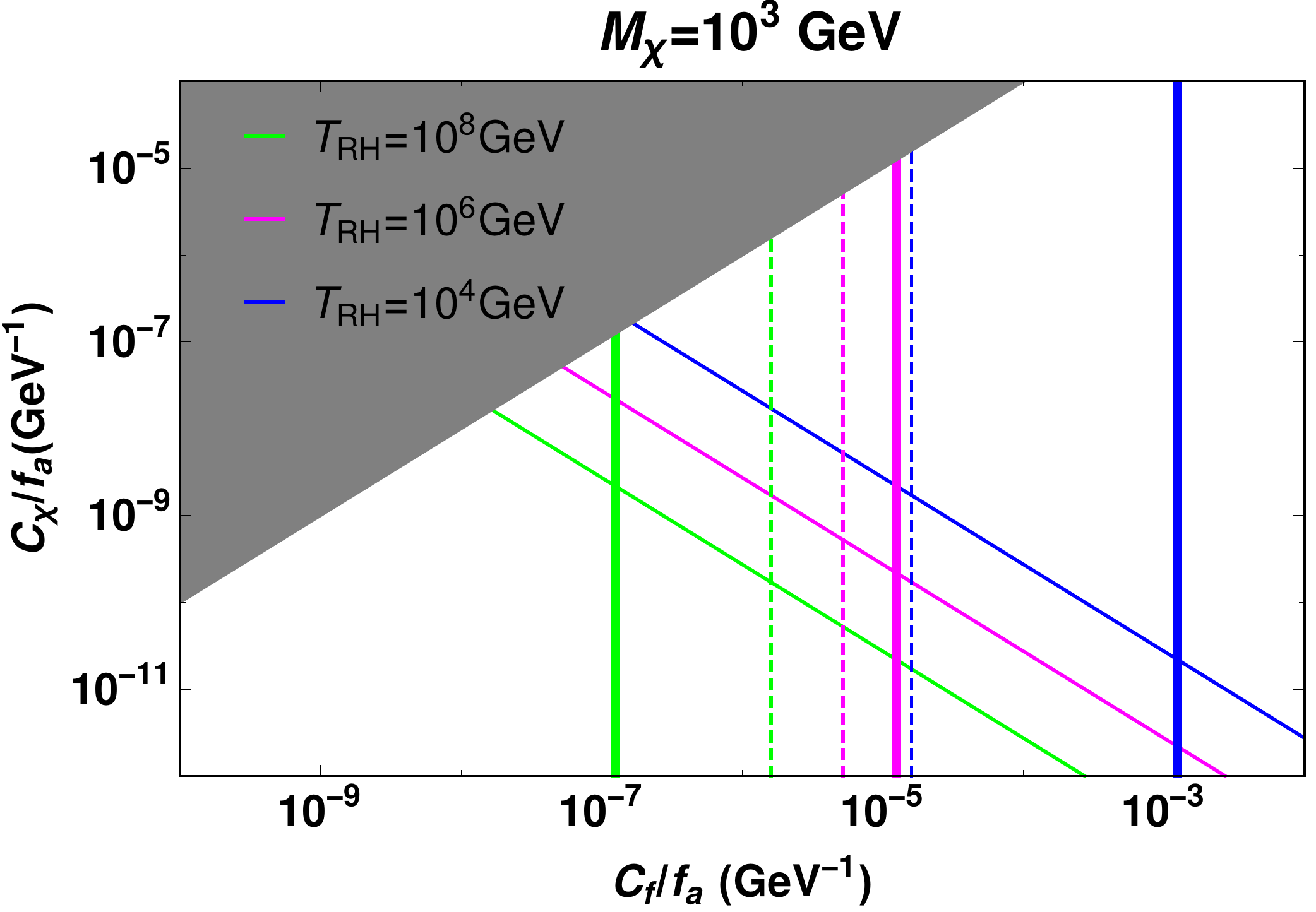}}
    \subfigure[\label{fig:f_th2}]{
    \includegraphics[scale=0.35]{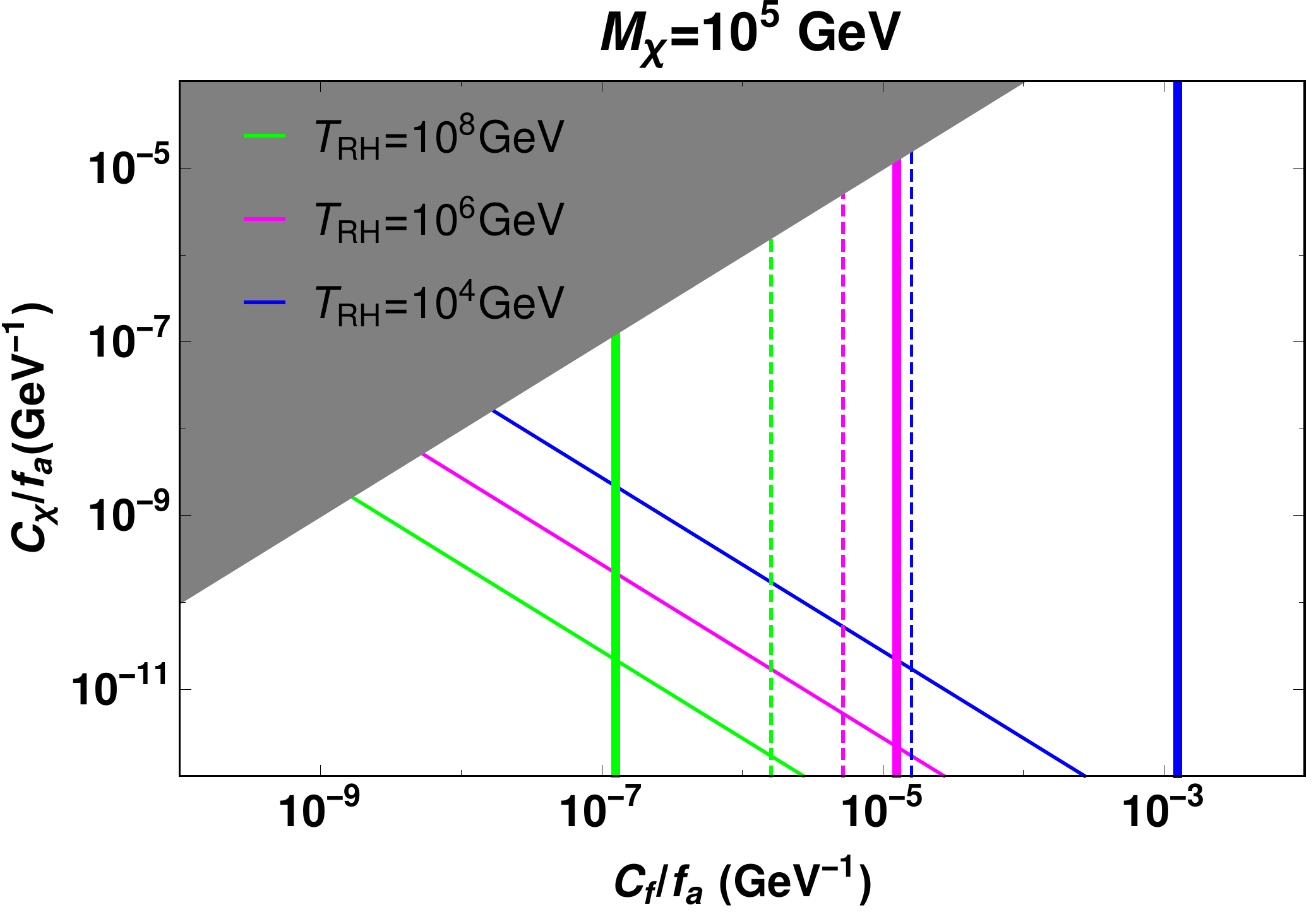}}
    \caption{ \it Allowed parameter space from non thermalization condition at early universe in $C_G/f_a~\g$ vs. $C_{\chi}/f_a~\g$ plane 
    shown in (a),(b){\bf (top panel)} and in $C_f/f_a~\g$ vs. $C_{\chi}/f_a~\g$ plane shown in (c),(d)({\bf (bottom panel)}. 
    We have chosen $M_{\chi}=10^3$ {\rm GeV} for the plots shown in (a),(c){\bf (left panel)} and $M_{\chi}=10^5$ {\rm GeV} for (b),(d ){\bf (right panel)}. The regions below different colored diagonal solid lines correspond to allowed parameter space where the DM production ($\text{SM SM}\to \chi \chi$) does not thermalize. The regions right to the colored dashed lines correspond to the parameter space where ALP production ($\text{SM SM}\to a a$) thermalizes. We have chosen a benchmark $M_{a}=0.1$ {\rm GeV} for all the plots.
     Different  values of  $T_{RH}\in {(10^8,10^6,10^4)}$ {\rm GeV} are shown by green, magenta, and blue respectively. The individual thick solid vertical lines correspond to the boundary of 
     the EFT valid zone for a given value of $T_{RH}$. The gray shaded region depicts where $C_\chi/f_a>C_i/f_a~(i\equiv G,f)$ and we restrict our analysis below that region for the reason explained in the text.}
    \label{fig:th}
\end{figure}
There is an upper bound on the effective ALP-gluon (fermion) coupling(s) and ALP-DM coupling from the (non-)thermalization conditions as described in eq.\eqref{eq:therm} or eq.\eqref{eq:thermf}. 
For sufficiently large couplings, processes like  $gg\to \chi \chi$ or $\Bar{f}f\to \chi \chi h$ may get thermalized (see fig. 3). 
The cross-sections for the processes $gg\to \chi \chi$ and $\Bar{f}f\to \chi \chi h$ are proportional to $\left(\frac{C_g}{f_a} \right)^{2} \left(\frac{C_{\chi}}{f_a} \right)^{2}$ and
$\left(\frac{C_f}{f_a} \right)^{2} \left(\frac{C_{\chi}}{f_a} \right)^{2}$ respectively (see eqs.\eqref{eq:therm1} or \eqref{eq:therm2}). 
 In Fig.\ref{fig:th} we present the (non-)thermalization constraint in  $C_G/f_a$ vs $C_{\chi}/f_a$ plane for gluon dominance  and
$C_f/f_a$ vs $C_{\chi}/f_a$ plane for fermion dominance. 
Within the wide range of allowed parameter space 
the relative strength between these effective couplings
($C_G/f_a,~ C_f/f_a$ and $C_{\chi}/f_a$) will decide the interplay of SM sector,
ALP and dark sector leading to numerous production mechanisms of DM as discussed in Ref.\cite{Hambye:2019dwd,Bharucha:2022lty}. 
However, we restrict our analysis in more simplified set up where we study only DM production from SM bath mediated by ALPs.
 ALPs produced from SM particles may get thermalized in early universe  leading to ALP freeze-out from thermal bath making the scenario more complicated.
For this reason we restrict our analysis to the parameter space where 
ALPs do not get thermalized.
This helps us to understand the underlying dynamics of the ALP portal
dark matter freeze-in process and enable us to probe the corresponding 
parameter space after taking into account constraints from 
various laboratory searches of ALPs.
To prevent the production of ALPs we impose  the thermalization condition on the processes $ij\leftrightarrow aa~ {\rm and}~i\leftrightarrow a a~(i,j=\text{SM particles})$ \cite{Bharucha:2022lty}.  
As large $C_G/f_a$ or $C_f/f_a$  leads to thermalization of ALPs irrespective of $C_{\chi}/f_a$, the bound is a vertical line in effective couplings plane further constraining the parameter space \cite{Hambye:2019dwd}.


We incorporate these conditions and the viable parameter space is shown in Fig.\ref{fig:th}.
In the top panel in Fig.\ref{fig:th1} and Fig.\ref{fig:th2} we consider gluon dominance with $M_{\chi}=10^3$ GeV (top left panel) and  $M_{\chi}=10^5$ GeV (top right panel) respectively. 
Parameter space using fermion dominance is shown in Fig.\ref{fig:f_th1} and Fig.\ref{fig:f_th2} with  $M_{\chi}=10^3$ GeV (bottom left panel) and  $M_{\chi}=10^5$ GeV (bottom right panel) respectively.
We set $M_a=0.1$ GeV for all four plots.
In grey shaded regions of all four aforementioned plots, $C_\chi/f_a> C_i/f_a~(i\equiv G,f)$ which are the required conditions for {\it sequential freeze-in} to happen. To prevent the occurrence of {\it sequential freeze-in} process for reasons discussed earlier, we confine our parameter space below that shaded region.
In Fig.\ref{fig:th} different colored lines in  blue, magenta and green   correspond to different values of $T_{RH}\in \{ 10^4,\,10^6 ,\, 10^8\}$ GeV respectively.  The area under the solid diagonal curves represent the allowed parameter space where DM production 
doesn't thermalize. 
The region correspond to the right of colored dashed lines corresponds to the parameter space where
ALPs thermalize.  
The thick solid vertical lines depict the upper limit on $C_i/f_a~(i\equiv G,f)$ for a fixed $T_{RH}$ without violating the EFT framework. While putting this upper limit, we consider the perturbative bound on individual Wilson coefficients $C_i \lesssim 4\pi$ and $T_{RH} < 4\pi f_a$ from the EFT limit. To satisfy this condition, we also confine our parameter space 
left to three coloured solid vertical lines corresponding to the aforementioned three fixed values of $T_{RH}$. 
From all plots in the above figure one notices that with increasing $T_{RH}$ one has to choose smaller couplings to meet the thermalization condition of DM production. The interaction strength  $\Gamma_{gg \to \chi \chi}$ and $\Gamma_{f \Bar{f} \to \chi \chi h}$ increases with temperature 
faster than the Hubble $H$, as dictated by the functional dependence on temperature (see eq.\eqref{eq:intg} and eq.\eqref{eq:intf}).
To compensate the increment in interaction strength  with higher $T_{RH}$, the effective couplings face more  stringent bound. 
Similarly, as the ALP production also increases faster than $H$ with increasing $T_{RH}$ the allowed value of $C_g/f_a$ ($C_f/f_a$) for gluon dominance (fermion dominance) is lower for higher $T_{RH}$. 
For this reason we see the stronger (weaker) limits on parameter space correspond to $T_{RH}=10^8$ GeV ($10^4$ GeV) shown by the area under the green (blue) lines for all four plots in Fig.\ref{fig:th}.
Also with increasing $M_{\chi}$ the interaction rate $\Gamma_{gg \to \chi \chi}$ increases. That is why we see lesser allowed parameter space in Fig.\ref{fig:th2} than in Fig.\ref{fig:th1} for a fixed $T_{RH}$.
Same physical reasoning holds also for fermion dominance which lead to 
less allowed parameter space in Fig.\ref{fig:f_th2} than in Fig.\ref{fig:f_th1} for same $T_{RH}$.
Another point we note from the above figure is that  
fermion dominance scenario results in larger allowed parameter space than in gluon dominance for a fixed 
$M_{\chi}$ and $T_{RH}$.
Comparing the plots in the left panels  we find that for same $T_{RH}$ the solid diagonal lines have moved to upward in bottom panel (fermion dominance) than in top panel (gluon dominance). 
The reason behind this feature is the color factor associated with gluon annihilation which leads to more interaction strength than fermion annihilation 
for same values of involved parameters. 
For the same reason the dashed lines have moved to rightward in bottom panel (fermion dominance) than in top panel (gluon dominance) for same $T_{RH}$.
All the effective couplings we use to evaluate relic density already lie within the allowed range. 

\begin{figure}[!tbh]
    \subfigure[\label{fig:rl1}]{
    \includegraphics[scale=0.35]{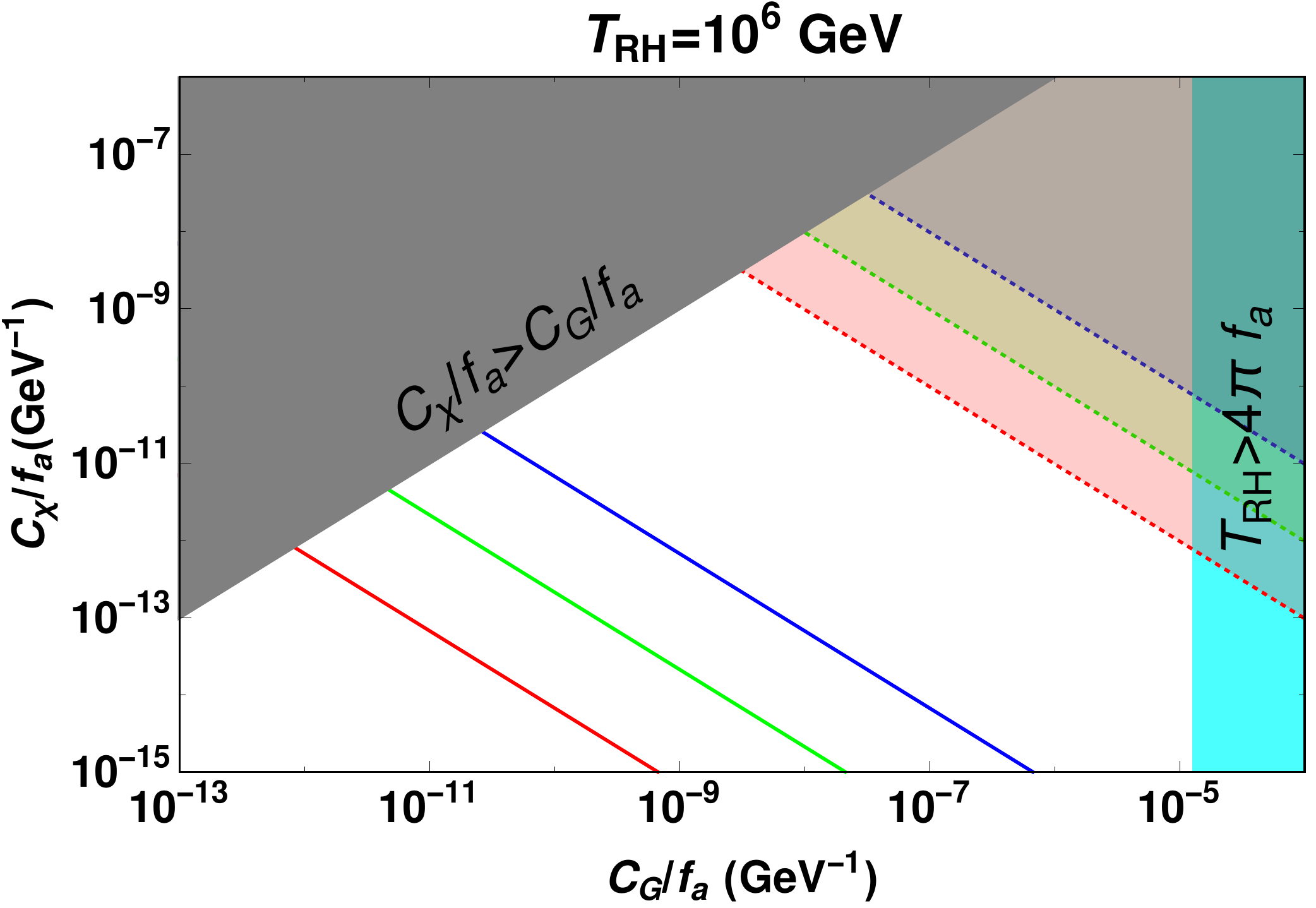}}
    \subfigure[\label{fig:rl2}]{
    \includegraphics[scale=0.35]{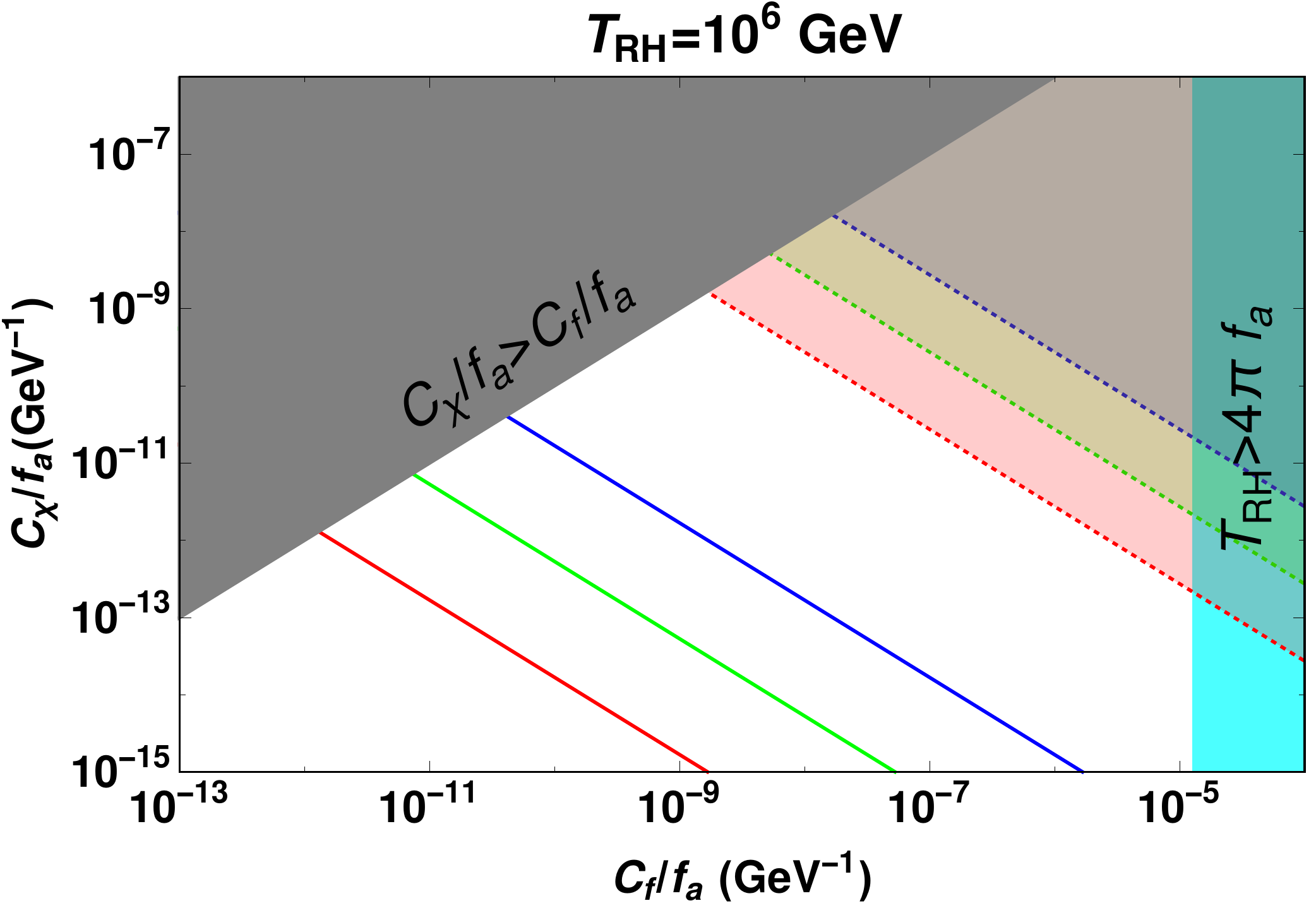}}
    \caption{ \it Observed relic density satisfying contours in $C_G/f_a~\g$ vs. $C_{\chi}/f_a~\g$ plane 
    shown in (a){\bf (left panel)} and in $C_f/f_a~\g$ vs. $C_{\chi}/f_a~\g$ plane shown in (b){\bf (right panel)}. 
    We have chosen a benchmark $M_{a}=0.1$ {\rm GeV} and $T_{RH}=10^6$ {\rm GeV} for both the plots. Relic density satisfying lines for different DM mass $M_{\chi}\in(10^5,10^4,10^3)$ {\rm GeV} shown in red,green and blue solid lines. The regions below different colored diagonal dotted lines correspond to allowed parameter space where the DM production ($\text{SM SM}\to \chi \chi$) does not thermalize. The gray shaded region depicts where $C_\chi/f_a>C_{G/f}/f_a$ and we restrict our analysis below that region for the reason explained in the text. The cyan shaded region depicts where $T_{RH}>4\pi f_a$ and EFT no longer holds at that region.}
    \label{fig:r}
\end{figure}

In Fig.\ref{fig:r} we represent the relic density satisfying contours in $C_i/f_a,~(i\equiv G,f)$ vs. $C_{\chi}/f_a$ plane. 
We consider gluon dominance in Fig.\ref{fig:rl1} and fermion dominance in Fig.\ref{fig:rl2} for a benchmark point $M_{a}=0.1$ GeV and  $T_{RH}=10^6$ GeV. 
In the gray shaded region  $C_\chi/f_a>C_{G/f}/f_a$ and in the light cyan shaded region $T_{RH}>4\pi f_a$.
Relic density satisfying lines for different DM mass $M_{\chi}\in(10^5,10^4,10^3)$ GeV shown in red,green and blue solid lines. 
The bound on the parameter space from thermalization condition is shown by  colored dotted lines corresponding to different values of $M_\chi$.
Thus our relic satisfying lines lie within the allowed parameter space.
The behaviour of the relic satisfying lines with $T_{RH}$ follows the same explanation as we elaborated in context of Fig.\ref{fig:th}.

We discuss the laboratory search prospects of our scenario in sec.\ref{sec:ALP_search}. However due to very tiny couplings most of the parameter space satisfying the observed relic is not within the reach of ALP search experiments as shown in Ref. \cite{Bharucha:2022lty}. This motivates us to explore the same scenario in the modified cosmological framework in the remainder part of this paper.

\medskip

\section{Non standard cosmology}
\label{sec:ns_cosmo}
In this work hitherto we assume the universe at the time of dark matter freeze-in 
production was radiation dominated (RD) whose energy density ($\rho_{rad}$) red-shifts with scale factor $a$ as $\rho_{rad}\sim a^{-4}$. The ALP portal couplings between 
the visible and dark 
sectors that provides the correct observed relic density is too small to be explored in current laboratory search for ALPs, as shown in  \cite{Kelly:2020dda,Bharucha:2022lty}. 
On the other hand to keep the DM and ALP in non-thermalized state (eq.\eqref{eq:therm} \& eq.\eqref{eq:thermf}) in the early universe, the effective coupling between the 
DM-pair and ALP must be small. This small ALP-portal coupling automatically leads to highly 
suppressed DM nucleon scattering cross-section, thus evading current
dark matter direct detection limits, which is the general feature of 
any freeze-in dark matter models.

The dark matter relic density highly sensitive to the cosmic evolution 
history. Any drastic change in the dynamics of the cosmic evolution may cause 
nontrivial impact on the dark matter relic density. So far, while discussing 
the dark matter relic density we have considered that the universe was 
radiation dominated (RD) at the time of BBN 
\cite{Kawasaki:2000en,Ichikawa:2005vw}. 
However, one cannot prohibit the possibility of some component other than
radiation dominating the total energy budget of the universe before BBN $(T \gg T_{\rm BBN})$.
Thus it is fair to assume that at early 
epoch of time, the universe was dominated by some non-standard species and the
corresponding scenario is popularly known as the non-standard (NS) 
cosmological framework \cite{ DEramo:2017gpl,DEramo:2017ecx,Barman:2022njh}.
The DM abundance in freeze-in scenario changes rather dramatically
if we assume cosmic history other than RD at the time of freeze-in 
\cite{DEramo:2017ecx,Calibbi:2021fld,Barman:2022njh}
which will be realized shortly at the end of this section. 
We present a brief summary  of NS cosmology here. Non standard cosmology 
may be realized by introducing a scalar field $\phi$ with equation of 
state (EOS) $p_{\phi}=\omega \rho_{\phi}$, where $p_{\phi}$, 
$\rho_{\phi}$ denote the pressure and energy density of
$\phi$ respectively and $\omega$ is the EOS 
parameter \cite{DEramo:2017gpl}. The energy density of such a species 
dilutes with scale factor $z$ as 
\begin{equation}
\rho_{\phi}\sim z^{-(4+n)},
\label{eq:rns}
\end{equation}
where, $n$ is given by $n= 3\omega-1 $.
For $n>0$ i.e. $\omega>\frac{1}{3}$, $\rho_{\phi}$ red-shifts faster than radiation energy density \cite{Ferreira:1997hj,Turner:1983he,Ferreira:1997au,Joyce:1996cp,Buchbinder:2007ad}.

In the early stage or at high temperature, in presence of this $\phi$ the total energy density ($\rho_{tot}$) of the universe emerges as  $\rho_{tot}= \rho_{rad}+ \rho_{\phi}$. 
As the universe cools down, it becomes radiation dominated again prior 
to BBN temperature $(T_{\rm BBN } \simeq 1{\rm~MeV})$~
\cite{Ichikawa:2005vw,Kawasaki:2000en}, so that the BBN 
predictions of light element abundances remain unaltered and during 
this process at some temperature $T_{eq} \gtrsim T_{\rm BBN}$, 
$\rho_{rad}$ becomes equal to $ \rho_{\phi}$. If one considers $T_{eq}$ 
close to $T_{\rm BBN} \simeq 1~{\rm MeV}$, the number of 
active neutrino degrees of freedom $(N_\nu)$ receives an additional contribution
from $\rho_{\phi}$ which can significantly modify bound on 
$\Delta N_{\nu} \equiv N_\nu - N^{\rm SM}_\nu$ from the 
BBN~\cite{Cyburt:2015mya}. Thus to comply
with this BBN bound $T_{eq}$ and $n$ are constrained by the
following condition\cite{ DEramo:2017gpl}:
\begin{equation}
T_{eq}\gtrsim (15.4)^{1/n}\,\text{MeV} 
\label{eq:tr}
\end{equation}
Using the definition of $T_{eq}$ as quoted in ref. \cite{ DEramo:2017gpl} and entropy conservation one can write the total energy density as,
\begin{equation}
\rho_{tot}(T)= \rho_{\rm {rad}}(T)\left[1+ \dfrac{g_{*}(T_{eq})}
{g_{*}(T)} \left(\dfrac{g_{*s}(T)}{g_{*s}(T_{eq})}\right)
^{\frac{4+n}{3}} \left(\frac{T}{T_{eq}} \right)^n \right],
\label{eq:fast}
\end{equation}
where, $g_{*}$ and $g_{*s}$ are the relativistic degrees of freedom associated with the energy density and entropy densities respectively.
With this total energy density in NS cosmology the Hubble parameter 
is defined as  $H(T)\sim \sqrt{\rho_{tot}}/ M_{pl}$ in contrast to RD universe where Hubble parameter scales as $H(T)\sim \sqrt{\rho_{rad}}/ M_{pl}$. 
Since for any $n>0$ and $T>T_{eq}$, $\rho_{tot}$ is larger than $\rho_{rad}$,$H(T)$ is also larger in NS cosmology compared to the case when the universe is only radiation dominated. Thus NS cosmology leads to the scenario when the universe expands faster than the usual RD universe.
At the limit $T\gg T_{eq}$ the Hubble scales as \cite{ DEramo:2017gpl} 
\begin{equation}
H(T)\sim \frac{T^2}{M_{pl}} \left(\frac{T}{T_{eq}} \right)^{n/2} 
\label{eq:hns}
\end{equation}

We will now study the phenomenological consequence of non standard cosmology in the context of our present set up and show how the DM phenomenology differs 
from that in RD universe for any $n>0$.

\begin{figure}[tbh!]
    \centering
    \includegraphics[scale=0.45
    ]{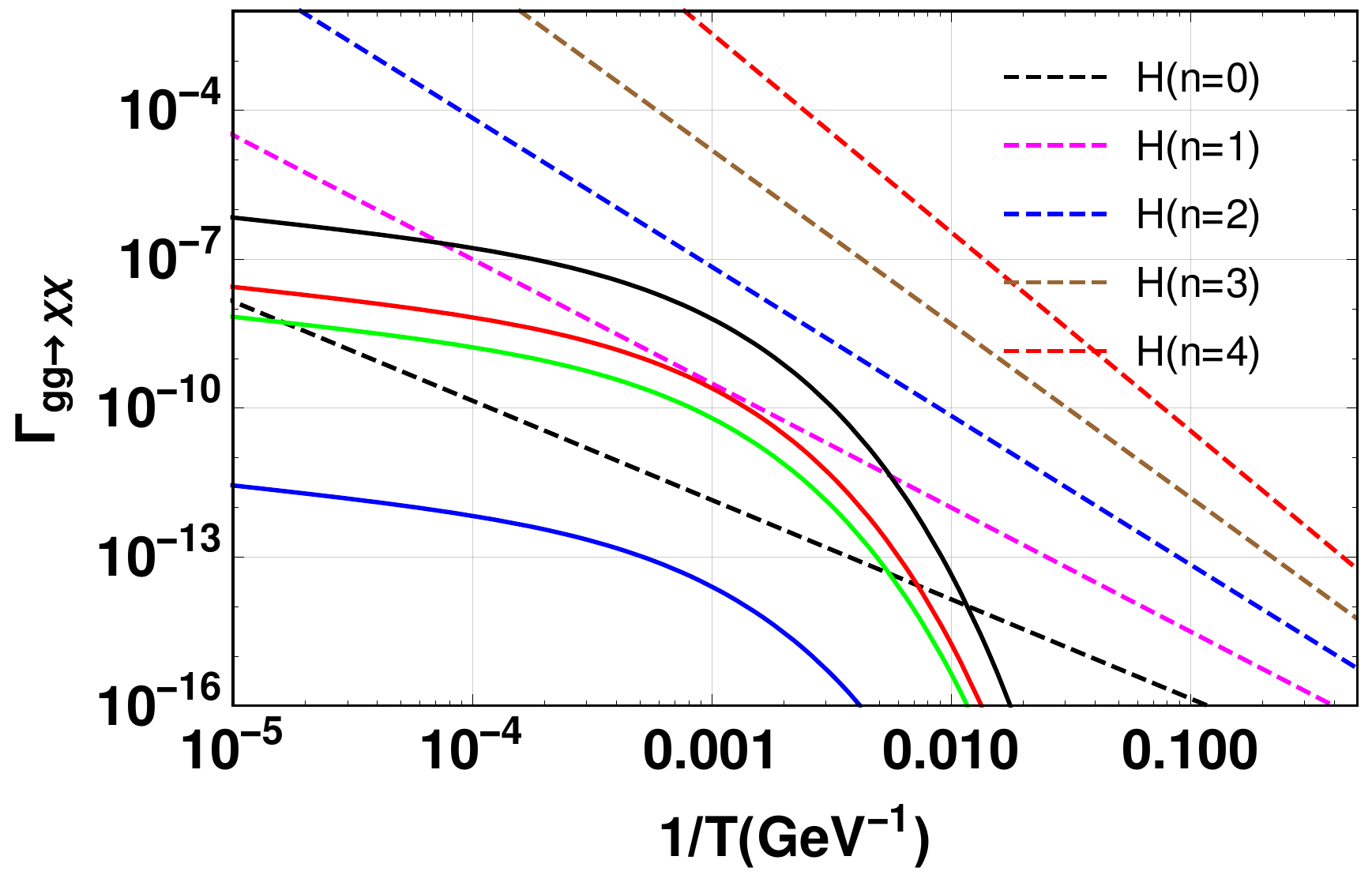}
    \caption{\it Evolution of $gg\to \chi \chi$ interaction rate ($\Gamma_{gg\to \chi \chi}$) with inverse of temperature ($1/T$) for $T_{RH}=10^5$ {\rm GeV}, $M_{\chi}=5\times 10^2$ {\rm GeV}, $C_{\chi}/f_a=10^{-8}\g$  and $T_{eq}=20$ MeV. The solid lines correspond to interaction rate for different values of $C_G/f_a \in {(\,10^{-7},\,10^{-6},\,5\times10^{-6},\,10^{-5})}${\rm GeV}$^{-1}$ as shown by blue, green, red and black lines respectively. The dashed lines correspond to $H$ for different $n\in\{0,1,2,3,4\}$ and depicted by black, magenta, blue, brown and red dashed lines respectively.  }
    \label{fig:non_stand}
\end{figure}
 After introducing the basic idea and motivation of the NS cosmology we now discuss the modified freeze-in abundance of DM in this framework.
 In such fast expanding universe, to keep the DM particle $\chi$ non-thermalized at early universe, eq.\eqref{eq:therm} has to be satisfied but with a modified Hubble parameter as in eq.\eqref{eq:hns}. 
 In  Fig.\ref{fig:non_stand} we show the evolution $\Gamma_{gg\to \chi \chi}$ with $1/T$ (GeV$^{-1}$) with 
 a benchmark value of $T_{RH}=10^5$ GeV, $T_{eq}=20$ MeV, $M_{\chi}=5\times10^2$ GeV and  $C_{\chi}/f_a=10^{-8}$GeV$^{-1}$  for four different values of $C_{G}/f_a$ depicted by 
 the solid lines ($C_{G}/f_a=10^{-7}$(Blue),$10^{-6}$(Green),$5\times10^{-6}$(Red),$10^{-5}$(Black) $\g$). 
 Whereas the dashed lines correspond to $H$ with different values of $n$ ($n=0$(Black),$n=1$(Magenta),$n=2$(Blue), $n=3$(Brown), $n=4$(Red).
Note that $n=0$ represents standard RD universe.
From Fig.\ref{fig:non_stand} we notice that with an increase in the values of $n$, $H(T)$ increases as elaborated before in the context of eq.\eqref{eq:fast}. 
From this figure it is also evident that with higher values $C_G/f_a$ ($\gtrsim 10^{-6}\g$), the process $gg\to \chi \chi$ gets thermalized in RD universe ($n=0$) but fails to thermalize in non-standard (NS) scenario ($n\neq0$). 
This leads to the relaxation of the parameter space in the effective coupling plane ($C_{g}/f_a$ vs. $C_{\chi}/f_a$), allowing larger values of these effective coupling for the non-thermal production of DM which will be analyzed in the next section. 
Similar kind of conclusion can be also drawn for fermion dominance also
i.e. for the evolution of $\Gamma_{\Bar{f}f\to \Bar{\chi}\chi}$. 

To understand the impact of this \textit{faster than radiation} expansion history of the universe on DM abundance we again solve eq.\eqref{eq:beq} with the modified 
Hubble parameter, $H$ defined in eq.\eqref{eq:hns}. The dark matter phenomenology in this framework is governed by the following free parameters:
\begin{equation}
C_g/f_a,~C_f/f_a,\,C_{\chi}/f_a,\,M_a,\,M_{\chi},\,T_{RH}, \, n, \, T_{eq}   
\end{equation}
where, $n$ and $T_{eq}$ are two additional free parameters representing the NS cosmology. 
\begin{figure}[tbh!]
    \centering
    \subfigure[\label{tr20}]{
    \includegraphics[scale=0.4]{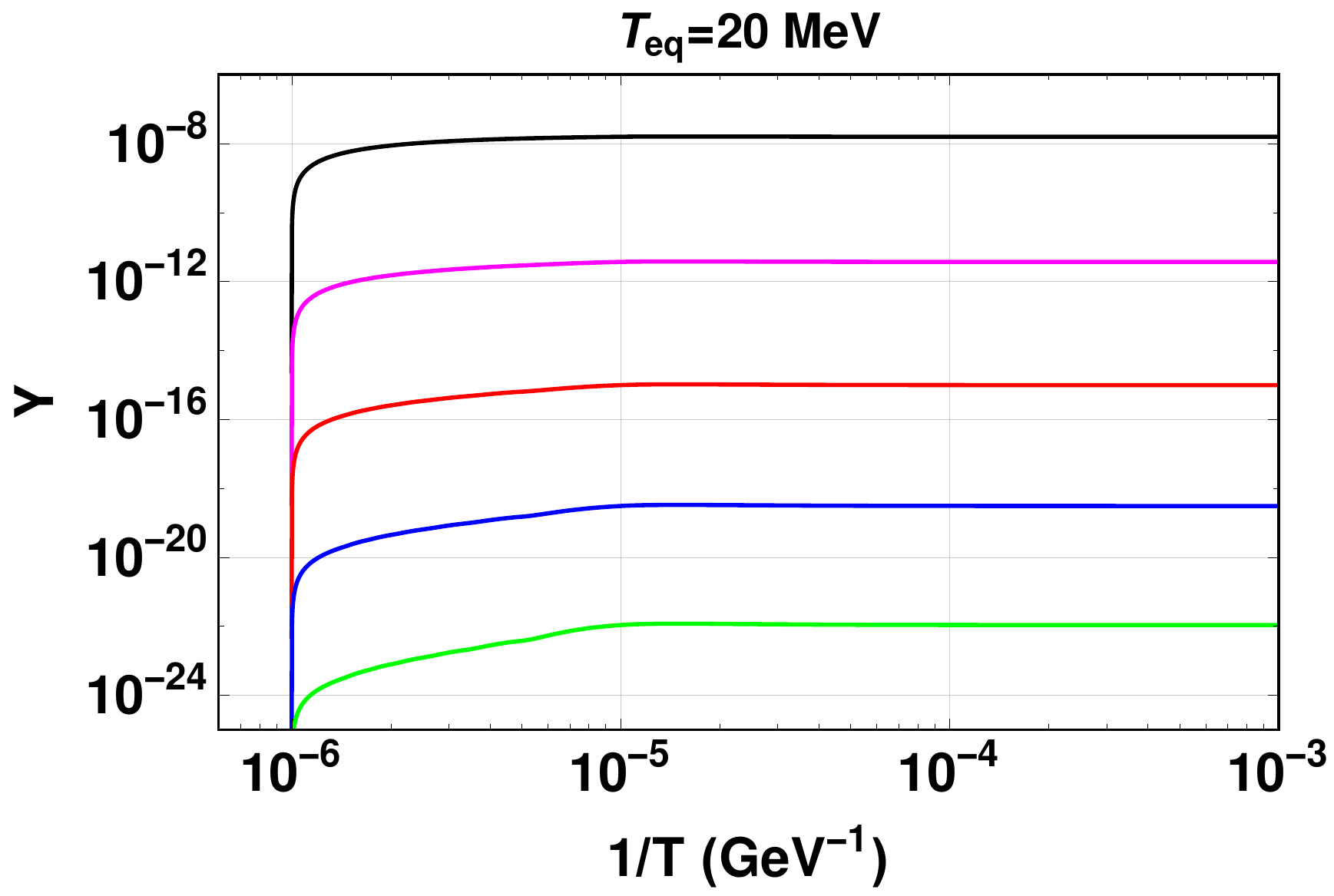}}
    \subfigure[\label{tr200}]{
    \includegraphics[scale=0.4]{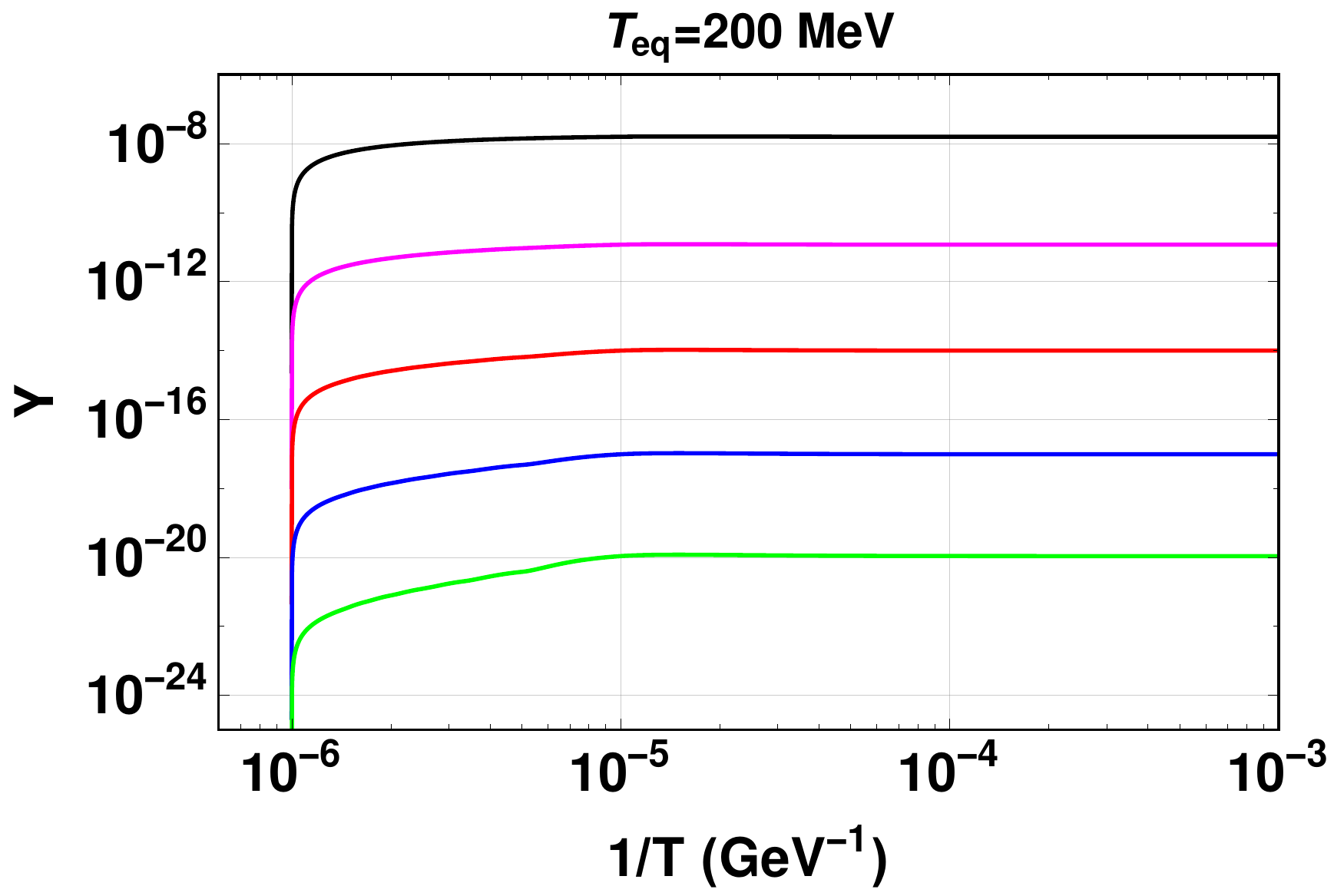}}
    \caption{\it Evolution of abundance $Y$ with $1/T$ for
     $T_{RH}=10^6$ {\rm GeV},
    $C_{g}/f_a=10^{-8}$ {\rm GeV}$^{-1}$, $C_{\chi}/f_a=10^{-10}\g$, 
    $M_{\chi}=10^3$ {\rm GeV}, $M_a=0.1$ {\rm GeV} with (a) $T_{eq}=20$ {\rm MeV} {\bf(left panel)} and (b) $T_{eq}=200$ {\rm MeV} {\bf(right panel)}. Different solid lines correspond to different values of $n\in (0,1,2,3,4)$ shown in black, magenta, red, blue, green colors respectively from top towards bottom. $n=0$ represents the standard cosmological scenario.}
    \label{fig:abund_ns}
\end{figure}
In NS cosmological scenario one has to be careful about the upper limit on $T_{RH}$ 
given by $T_{RH}\lesssim M_{pl} (T_{eq}/M_{pl})^{\frac{n}{n+4}}$ as pointed out in ref.\cite{DEramo:2017gpl}. 
While doing our analysis in NS cosmology we ensure that for all model parameters, this inequality holds.
Considering the gluon dominance process
we solve eq.\eqref{eq:beq} to evaluate the co-moving abundance
$Y$ with different values of $n\in (0,1,2,3,4)$ and present the evolution of $Y$ with $1/T$ (GeV$^{-1}$) 
for $T_{RH}=10^6$ GeV, $C_{G}/f_a=10^{-8}$ GeV$^{-1}$, $C_{\chi}/f_a=10^{-10}$ GeV$^{-1}$, 
$M_{\chi}=10^3$ GeV and $T_{eq} = 20 (200)$ (MeV) in Fig. \ref{tr20}(\ref{tr200}).
The different solid lines correspond to different values of $n\in(0,1,2,3,4)$ are depicted by black, magenta, red, blue, green colours respectively.
Based on the aforementioned plot, it is clear that for a fixed value of $T_{ef}$, an increase in the value of $n$ leads to a decrease in the value of $Y$.
Eq. \eqref{eq:hns} shows that for a fixed value of $T_{eq}$, an increase in the value of $n$ leads to faster expansion rate denoted as $H$. This faster expansion rate of the universe results in the suppression of non thermal DM 
abundance as $Y\sim 1/H$. 
This feature is highly luminous in both the plots in Fig.\ref{tr20} $\&$ Fig.\ref{tr200}. Comparing
Fig.\ref{tr20} and Fig.\ref{tr200} we perceive that for same $n>0$ with higher $T_{eq}$ leads to higher DM abundance. 
This can be readily comprehended by noting the functional dependence of $H$ on $T_{eq}$ in eq.\eqref{eq:hns} which apprises that an increase in $T_{eq}$ leads to a decrease in $H$ consequently an increase in $Y$.  
In the aforementioned figure, the line corresponding to $n=0$(black line) represents the RD universe which remains unaltered with the variation of $T_{eq}$.
This overall picture is also valid in the case of fermion dominance production of DM.

From the above discussion  we conclude that same couplings will produce lower abundance in the fast expanding (NS) universe than in RD universe. Needless to say while 
other parameters are kept fixed, one has to pick higher values of effective couplings in fast expanding scenario than in RD universe in order to satisfy the 
correct relic density without violating the thermalization bound.
Such large value of effective ALP couplings can be further probed using various laboratory searches of ALP.
This motivates us to  explore the ALP phenomenology in the non standard cosmological scenario in the rest of this paper. 

\section{Relic Density in Non-standard cosmology}
\label{sec:relic_ns}

Following the discussion of the previous section here we investigate the consequences of non-standard cosmology in deciding the relic density. 
\begin{figure}[tbh]
    \centering
    \subfigure[\label{fig:nsa}]{
    \includegraphics[scale=0.35]{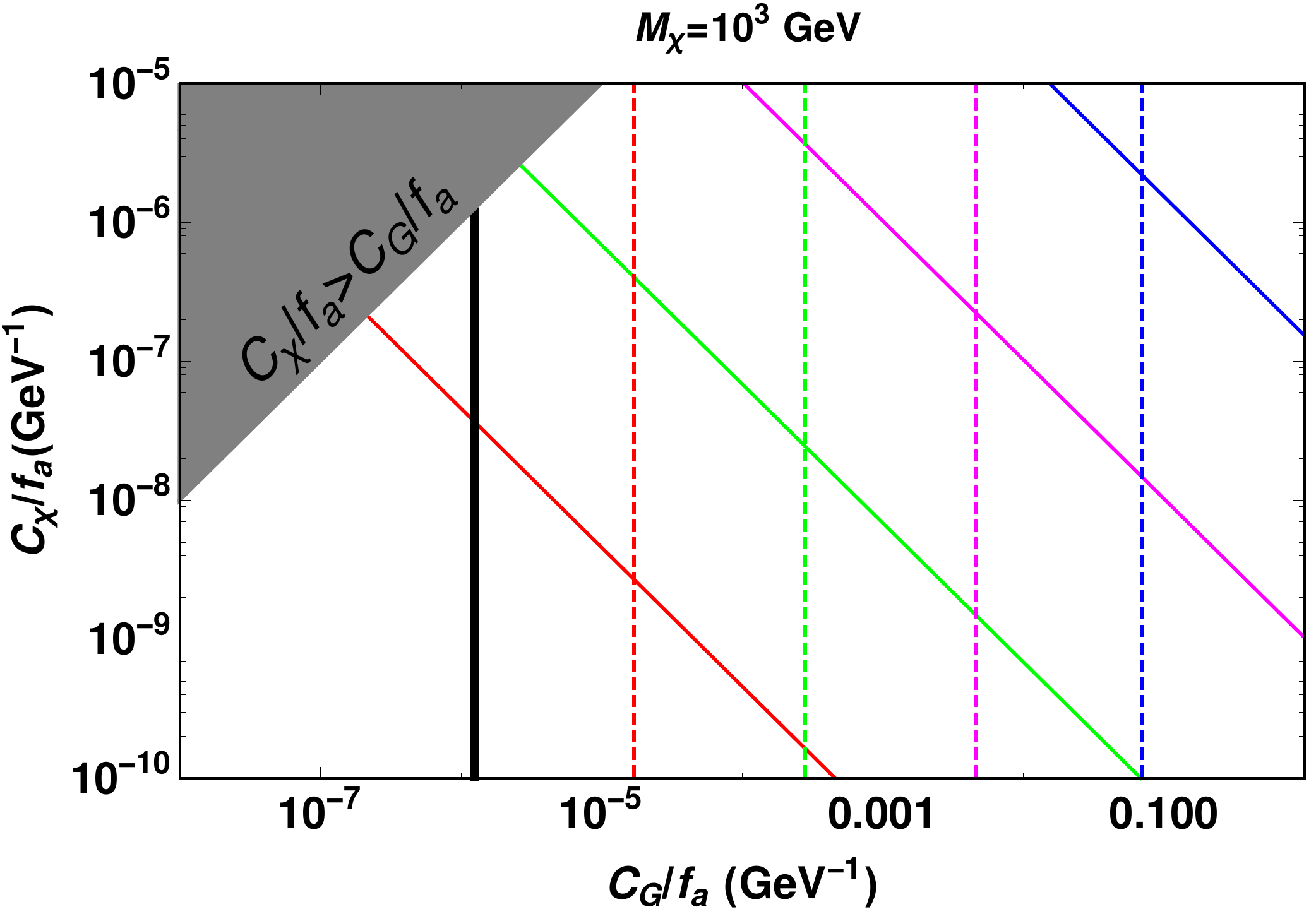}}
    \subfigure[\label{fig:nsb}]{
    \includegraphics[scale=0.35]{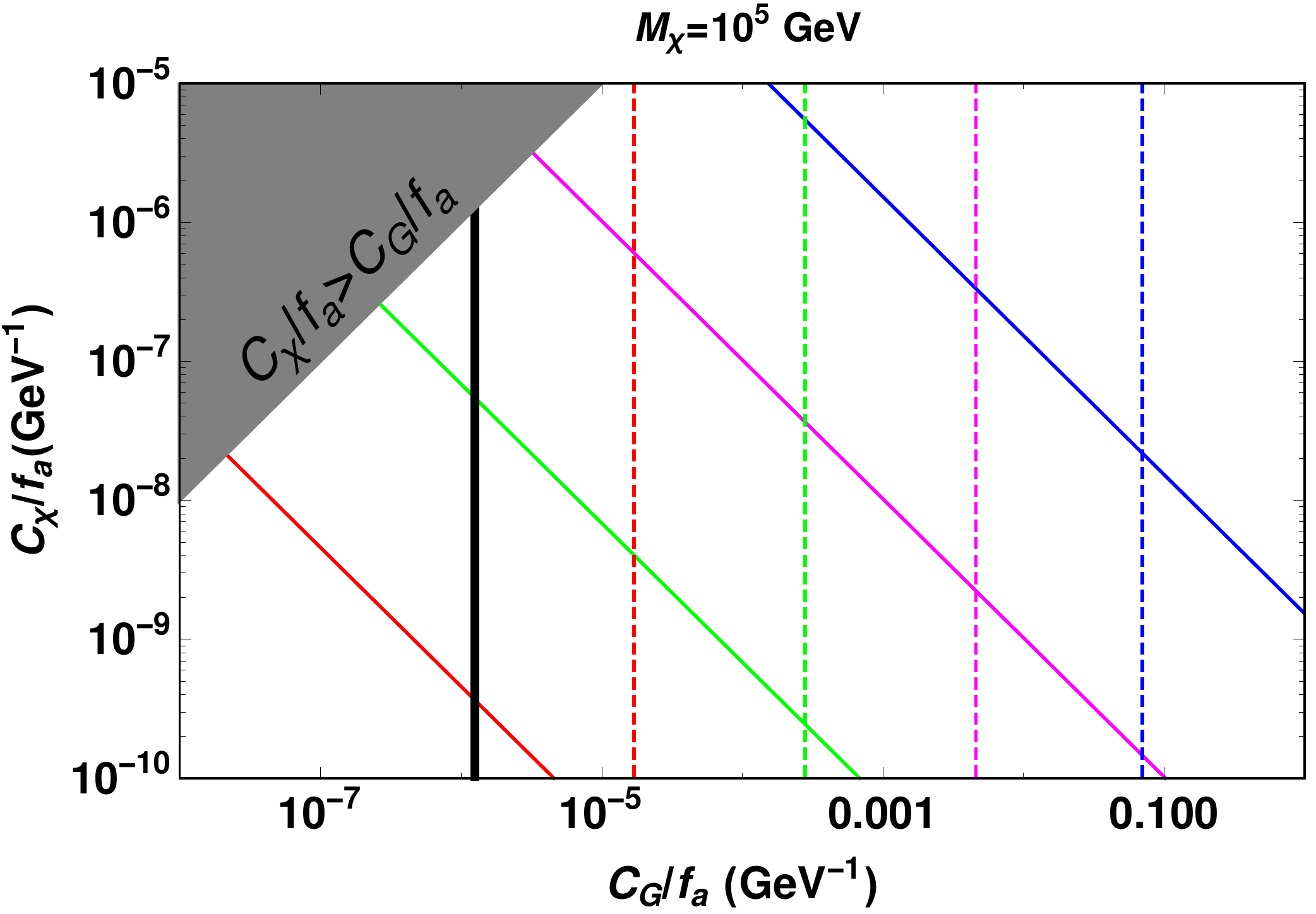}}
    \subfigure[\label{fig:fnsa}]{
    \includegraphics[scale=0.35]{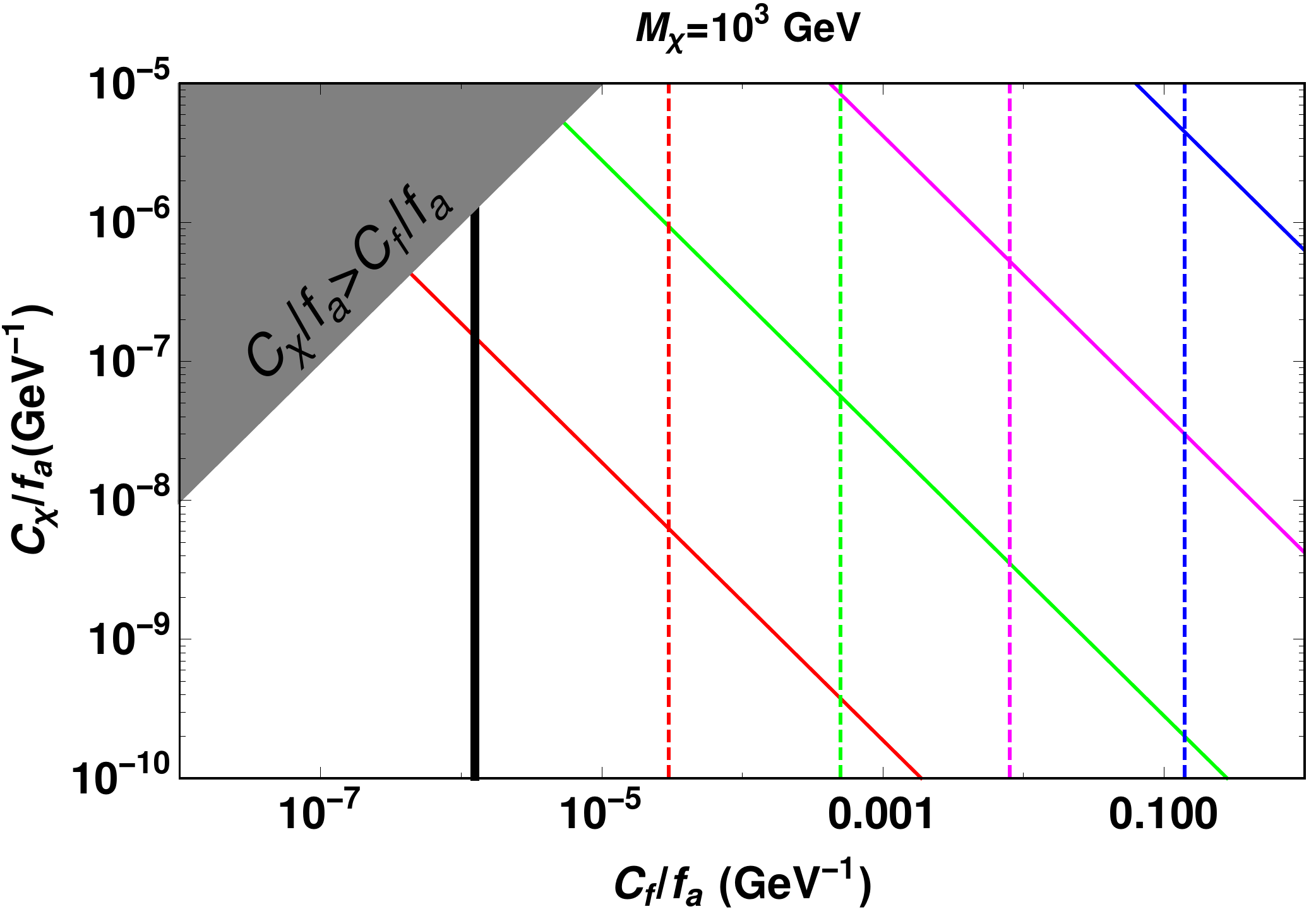}}
    \subfigure[\label{fig:fnsb}]{
    \includegraphics[scale=0.35]{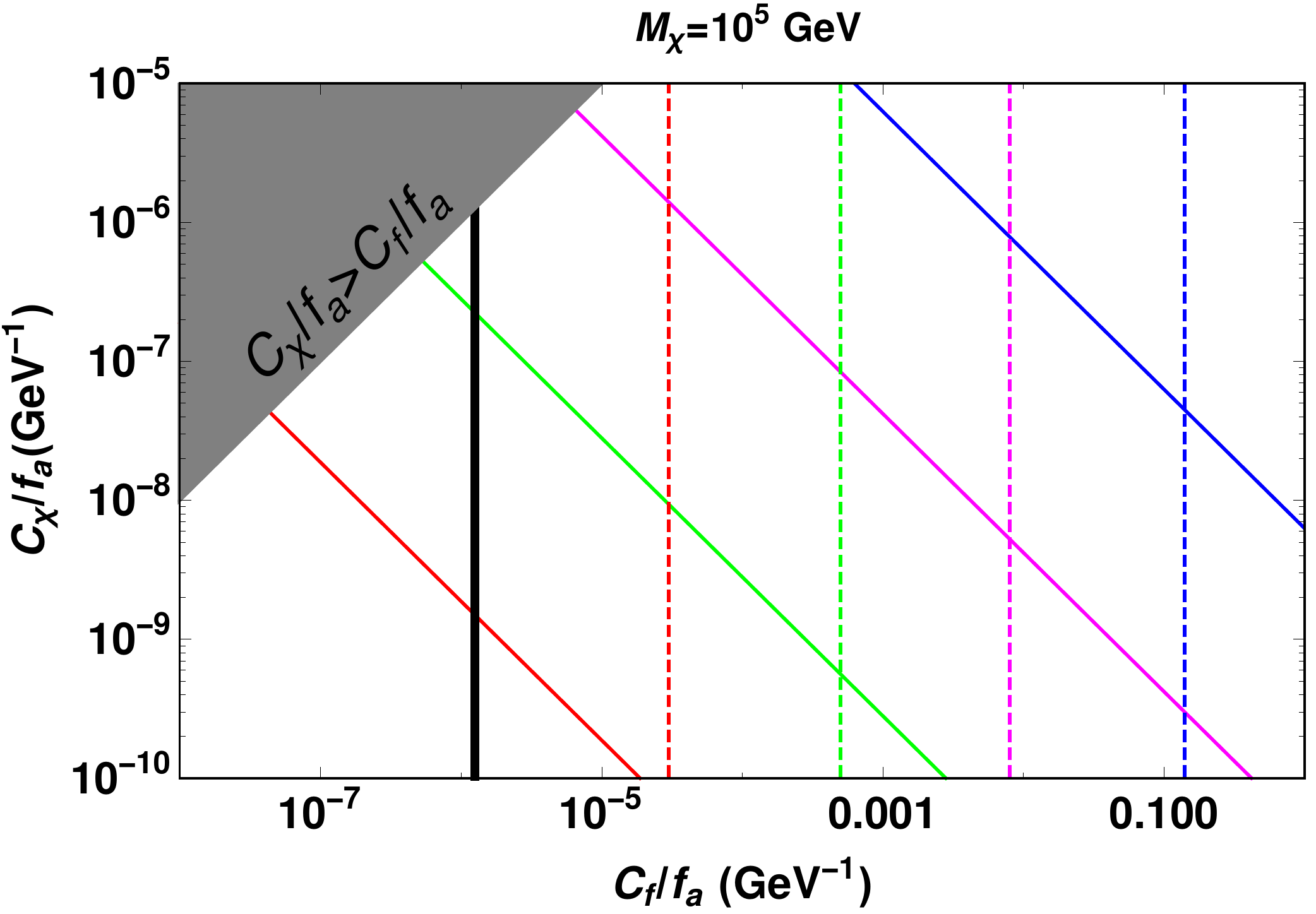}}
    \caption{\it Allowed parameter space from (non-)thermalization condition at early universe in $C_G/f_a~(\g)$ vs. $C_{\chi}/f_a~(\g)$ plane 
    shown in (a),(b){\bf (top panel)} and in $C_f/f_a~(\g)$ vs. $C_{\chi}/f_a~(\g)$ plane shown in (c),(d)({\bf (bottom panel)}. 
    We have chosen $M_{\chi}=10^3$ {\rm GeV} for the plots shown in (a),(c){\bf (left panel)} and $M_{\chi}=10^5$ {\rm GeV} for (b),(d ){\bf (right panel)}. We have chosen a benchmark $M_{a}=0.1$ {\rm GeV}, 
    $T_{RH}=10^7$ {\rm GeV} and $T_{eq}=20$ {\rm MeV} for all the plots.
    The regions below different colored solid lines correspond to allowed parameter space where the DM production ($\text{SM SM}\to \chi \chi$) does not thermalize. The regions right to the colored dashed lines correspond to the parameter space where ALP production ($\text{SM SM}\to a a$) thermalizes.
    The different colored curves (both solid and dashed) correspond to $n\in\{1,2,3,4\}$as shown by red,green,magenta and blue respectively.
    The  thick solid vertical line corresponds to the boundary of 
     the EFT valid zone for the given value of $T_{RH}$. The gray shaded region depicts where $C_\chi/f_a>C_i/f_a~(i\equiv G,f)$ and we restrict our analysis below that region for the reason explained in the text.
   }
    \label{fig:therm_ns}
\end{figure}
In Fig.\ref{fig:therm_ns} we present the allowed parameter space where the DM production processes do not thermalize at the time of freeze-in.  
We show the parameter space in  $C_G/f_a$ vs $C_{\chi}/f_a$ plane for gluon dominance  and
$C_f/f_a$ vs $C_{\chi}/f_a$ plane for fermion dominance. 
In  Fig.\ref{fig:nsa} and Fig.\ref{fig:nsb} we consider gluon dominance for $M_{\chi}=10^3$ GeV and  $M_{\chi}=10^5$ GeV respectively.
Parameter space using fermion dominance is shown in Fig.\ref{fig:fnsa} and Fig.\ref{fig:fnsb} for $M_{\chi}=10^3$ GeV  and $M_{\chi}=10^5$ GeV respectively.
All the above four plots are generated assuming $M_a=0.1$ GeV,$T_{eq}=20$ MeV and $T_{RH}=10^{7}$ GeV.
The regions left to solid coloured diagonal lines correspond to the allowed parameter space in the effective coupling plane where DM production 
doesn't thermalize. 
The region correspond to the right of colored dashed lines represents the parameter space where
ALPs thermalize. 
The red, green, magenta and blue lines (both solid and dashed)  correspond to different values of $n=1,2,3,4$  respectively.

From all four plots in Fig.\ref{fig:therm_ns} we notice that with increase in $n$, the allowed parameter space also increases in the effective coupling plane. 
This feature can be explained using eq.\eqref{eq:hns} where we see the expansion rate of universe  $H$ increases 
with increase in $n$ and with this larger value of $H$ both the DM production process fails to thermalize 
 even with higher values of $C_G/f_a$ (or,  $C_f/f_a$)and $C_{\chi}/f_a$
 satisfying eqs. \eqref{eq:therm} and \eqref{eq:thermf}. 
Comparing plots in left panel (Fig.\ref{fig:nsa},\ref{fig:fnsa})
and right panel (Fig.\ref{fig:nsb},\ref{fig:fnsb}) one notices that larger parameter space is allowable 
for lower values of $M_{\chi}$. This is because
the interaction rate is proportional to $M_{\chi}^2$ (eq. \eqref{eq:cf3}), bearing that smaller $M_{\chi}$ results 
lower interaction rates which satisfy the thermalization conditions shown in eqs.\eqref{eq:therm} and \eqref{eq:thermf}.
Comparing Fig.\ref{fig:th} and Fig.\ref{fig:therm_ns} one observes that for a same $T_{RH}$ non standard cosmology provides larger values of effective couplings within the allowed parameter space in effective coupling plane than in the standard cosmology (RD). 
This behaviour is compatible with what we discussed in the previous section. This opens up the opportunity to use higher effective couplings in DM production without violating the conditions shown in eqs.\eqref{eq:therm} and \eqref{eq:thermf}. 
as demonstrated in above figure.


\begin{figure}[tbh]
    \centering
    \includegraphics[scale=0.5]{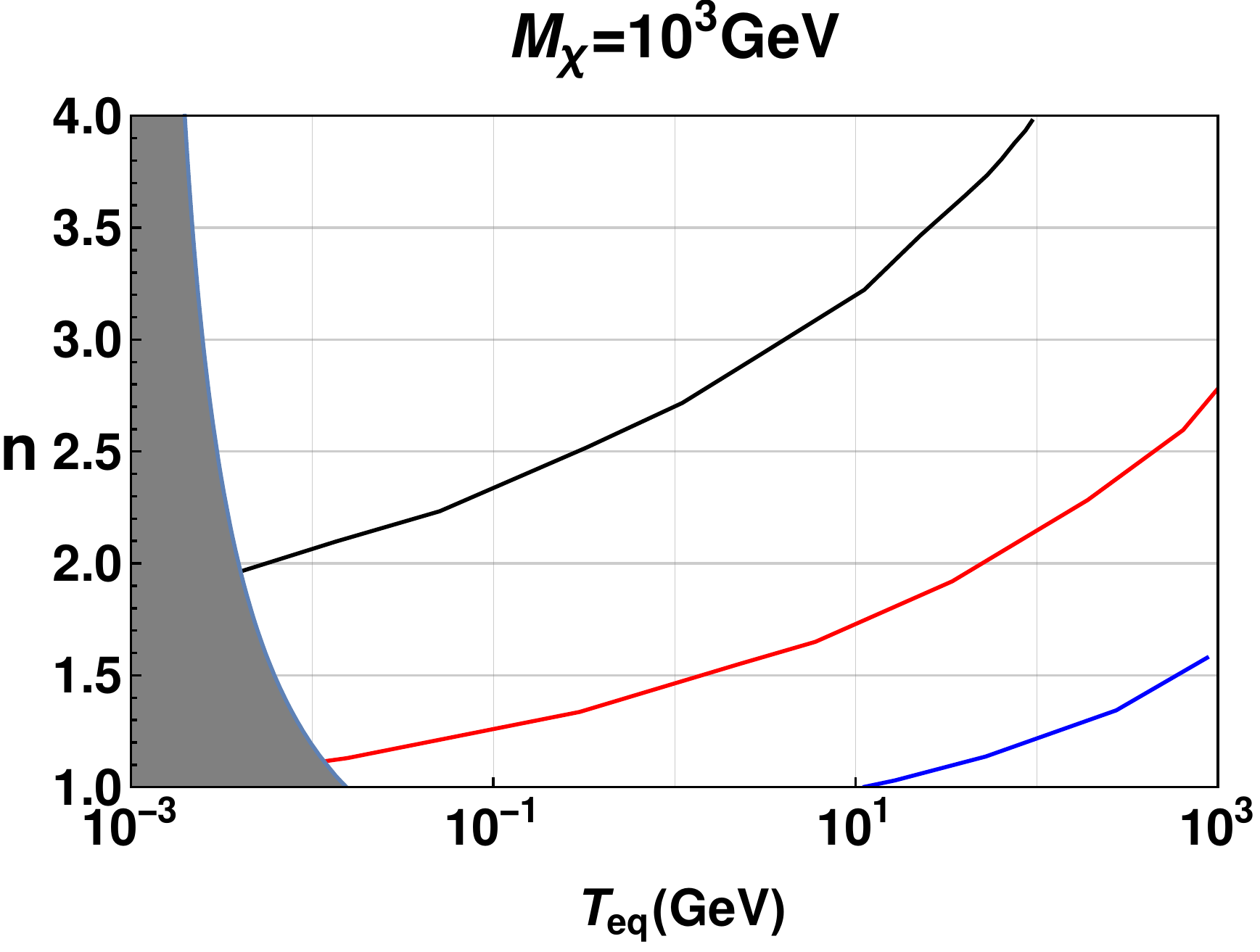}
    \caption{\it Lines satisfying the  observed relic abundance for different
values of $C_G/f_a$ for  $T_{RH}=10^7$ {\rm GeV}, $C_{\chi}/f_a=10^{-10}\g,\, M_{\chi}=10^3$ {\rm GeV} in the $T_{eq}$ vs. $n$ plane. 
The grey shaded region is excluded from the BBN constraint on $T_{eq}$.
$C_G/f_a$ has been fixed to  $4\times 10^{-9}$(blue), $4\times 10^{-8}$(red), $4\times 10^{-6}$(black) $\g$.}
\label{fig:n_Tr}
\end{figure}

In Fig.\ref{fig:n_Tr} we present the relic satisfied contours in  $T_{eq}$ vs. $n$ plane to showcase the dependence of $n$ and $T_{eq}$ in deciding relic density while other parameters are kept fixed. 
Here we consider gluon dominance and take fixed values of   $T_{RH}=10^7$ GeV, $C_{\chi}/f_a=10^{-10}\g,\, M_{\chi}=10^3$ GeV and the different colored lines depict different values of $C_{G}/f_a\in (4\times10^{-9},\,4\times10^{-8},\,4\times10^{-6})\g$ shown in magenta, blue, red and black curves respectively.
The gray shaded region is excluded from the bound on $T_{eq}$ from BBN as mentioned in eq.\eqref{eq:tr}. 
One can see that for a fixed $T_{eq}$ higher  $C_G/f_a$ demands higher values of $n$  to satisfy the relic density as we have already pointed out in the context of Fig.\ref{fig:abund_ns}.
Hence we see $C_G/f_a=4\times10^{-6}~\g$ (black line) contour lies on the top while $C_G/f_a=4\times10^{-9}~\g$ (blue line) in the bottom in $T_{eq}-n$ plane. 
Another noteworthy feature from the above mentioned figure is that for a fixed value of $C_g/f_a$ an increase in $T_{eq}$ calls for an increase in $n$ also. This can be explained by the fact that  higher $T_{eq}$ results lower value of $H$ and to satisfy the relic density $n$ has to be increased to compensate the effect of $T_{eq}$.

\section{Laboratory and Astrophysical searches}
\label{sec:ALP_search}
In this section we will discuss the direct detection prospects of ALP in our set up. We present all sources of astrophysical, cosmological and collider constraints depending on the mass of ALP and effective ALP-SM couplings. 
In non-standard (NS) cosmological scenario, our results suggest that there is a
wide range of ALP-SM and ALP-DM effective couplings that 
satisfies the observed relic density. However, a large portion of this 
parameter space, especially the region with stronger ALP-SM effective 
couplings, is ruled out by the constraints from direct ALP searches.
In contrast, in  radiation-dominated (RD) universe, there are parameter 
regions that agree with the observed dark matter relic density but are 
still below the detection sensitivity of direct ALP searches. This is 
because the involved ALP-SM effective couplings are weaker in the standard cosmological (or RD) scenario.
We now discuss implications of these constraints on our scenario.
\begin{itemize}
    \item { \bf{Astrophysical constraints:}}
    Among the astrophysical bounds SN1987A puts the most stringent bound 
    on the effective ALP-SM coupling. 
    In standard picture of core-collapse supernova, it cools by emitting most of its energy via neutrino emission. However it might get affected in the presence of additional weakly interacting particles which can carry away additional energy from the core. 
    Likewise in our set up  ALP-gluon couplings ($C_G/f_a$) or, ALP-fermion couplings ($C_f/f_a$) 
    will be responsible for additional cooling via  nucleon ($N$) scattering ($N\,N\to N\,N\,a$) \cite{Ertas:2020xcc,Raffelt:1990yz,Chang:2018rso}.
   For a large ALP couplings although ALP would be produced in a large number due to stronger interaction, the same coupling will lead to re-absorption of  ALPs in the core \cite{Ertas:2020xcc}.
    For small ALP mass ($M_a<100$ MeV) the constraint from SN1987A
    is relevant and we incorporate that bound in our analysis
    \cite{Ertas:2020xcc,Kelly:2020dda}.  For other constraints from supernovae, see Refs. \cite{Hoof:2022xbe,Ferreira:2022xlw}.
     There is another relevant constraint coming from Horizontal Branch (HB) star \cite{Pospelov:2008jk,Avignone:1986vm} and global analysis in general \cite{Balazs:2022tjl}.  
    Any pseudo-scalar that can couple to electron or photon can interact
 with particles in the core of HB stars  and disrupt the stellar evolution. 
 Similar to SN1987A , ALPs can also affect the cooling mechanism of HB stars and hence constraints from HB stars put limit on the effective coupling of ALPs \cite{Bharucha:2022lty}.
 However this bound is sensitive only in the keV range of $M_a$ \cite{Pospelov:2008jk}. 
    \item{\bf Cosmological constraints:} For sufficiently low $M_a$ values, ALP 
    may be relativistic even at the time of BBN.
    The presence of additional relativistic pseudo-scalar degrees of freedom would naturally contribute to the radiation energy density and thus increase the expansion rate of universe $H$. Faster expansion of the universe can alter the light element abundances at the time of BBN. Thus BBN puts an upper bound on the effective number of degrees of freedom ($\Delta N_{\rm eff}$) and this can be used to constrain the ALP mass vs. coupling plane \cite{Millea:2015qra,Depta:2020wmr}. 
    \item{\bf Collider searches:}
    In the presence of ALP and SM effective couplings, ALP can decay into a pair of photons via loop for 
    sufficiently low value of $M_a$. However, for higher value of $M_a$, the hadronic decay channels open up.
    Electron beam dump experiment (E137) puts constraint on the effective ALP-SM coupling \cite{Bjorken:1988as}. 
    Here ALP can be produced by bremsstrahlung or $e^+e^-$ annihilation and decay to visible particles ($a\to e^+e^-,\gamma \gamma$) via either tree level or loop level processes.
    On the other hand for small effective coupling with SM, ALP becomes a long lived particle (LLP) which can be separable from background \cite{Hook:2019qoh}. These LLP ALP can be produced in p-p collision at the LHC and can be within 
    the reach of displaced decay search with the LHC Track trigger \cite{Hook:2019qoh,Kelly:2020dda}. 
    The FASER experiment at LHC will also be able to probe ALPs with mass above a few MeV \cite{FASER:2018eoc}.
    Another bound on ALPs come from the decay of heavy mesons and this 
    bound is sensitive in $M_a\sim$ MeV to a few GeV range. 
    For processes like $B^+\to K^+ a(\to \mu^+ \mu^- / e^+ e^-)$ 
    LHCb puts limit on effective ALP-SM coupling \cite{LHCb:2016awg}. 
    Experiment like CHARM also constrains the effective coupling
    considering the process $B^{0}\to K^{*0} a(\to \mu^+ \mu^- / e^+ e^-)$
    \cite{Dobrich:2018jyi}.
    For MeV scale mass, another constraint arises from NA62 which is designed to measure rare Kaon decays. 
    NA62  puts constraint on the effective ALP-SM coupling using
    the decay of charged Kaon, $K^+ \to \pi^+ + a $, 
    where ALP, $a$ acts as the long lived particle \cite{NA62:2021zjw,Ertas:2020xcc}. 
    This experiment is sensitive when $M_a<M_{\pi}$, however, as $M_a \approx M_\pi$, 
    the background becomes too large to have any meaningful limit on the ALP parameter space. The DUNE (Deep Underground Neutrino Experiment) can also be used to search for ALPs where ALPs are produced 
    with meson mixing or with gluon fusion \cite{Bauer:2017ris} and later 
    they decay to photons  or hadrons in the near detector (ND) \cite{Kelly:2020dda}.
    
\end{itemize}
After demonstrating all kinds of existing constraints and future sensitvity reaches we present the result in $M_a$ vs. effective coupling plane in the following two subsections
\footnote{For other relevant ALP constraints, see Refs. \cite{Cadamuro:2011fd,Langhoff:2022bij}}.

\subsection{Gluon dominance scenario}
Here, we consider gluon dominance scenario where only $C_G/f_a$ and $C_{\chi}/f_a$ are non zero while rest of the effective ALP couplings are zero.
We now present all the aforementioned ALP direct search constraints as well as the DM relic density satisfying points in the $M_a$ vs. $C_G/f_a$ plane.

\begin{figure}[tbh!]
    \centering
    \subfigure[\label{ls_g_1}]{
    \includegraphics[scale=0.52]{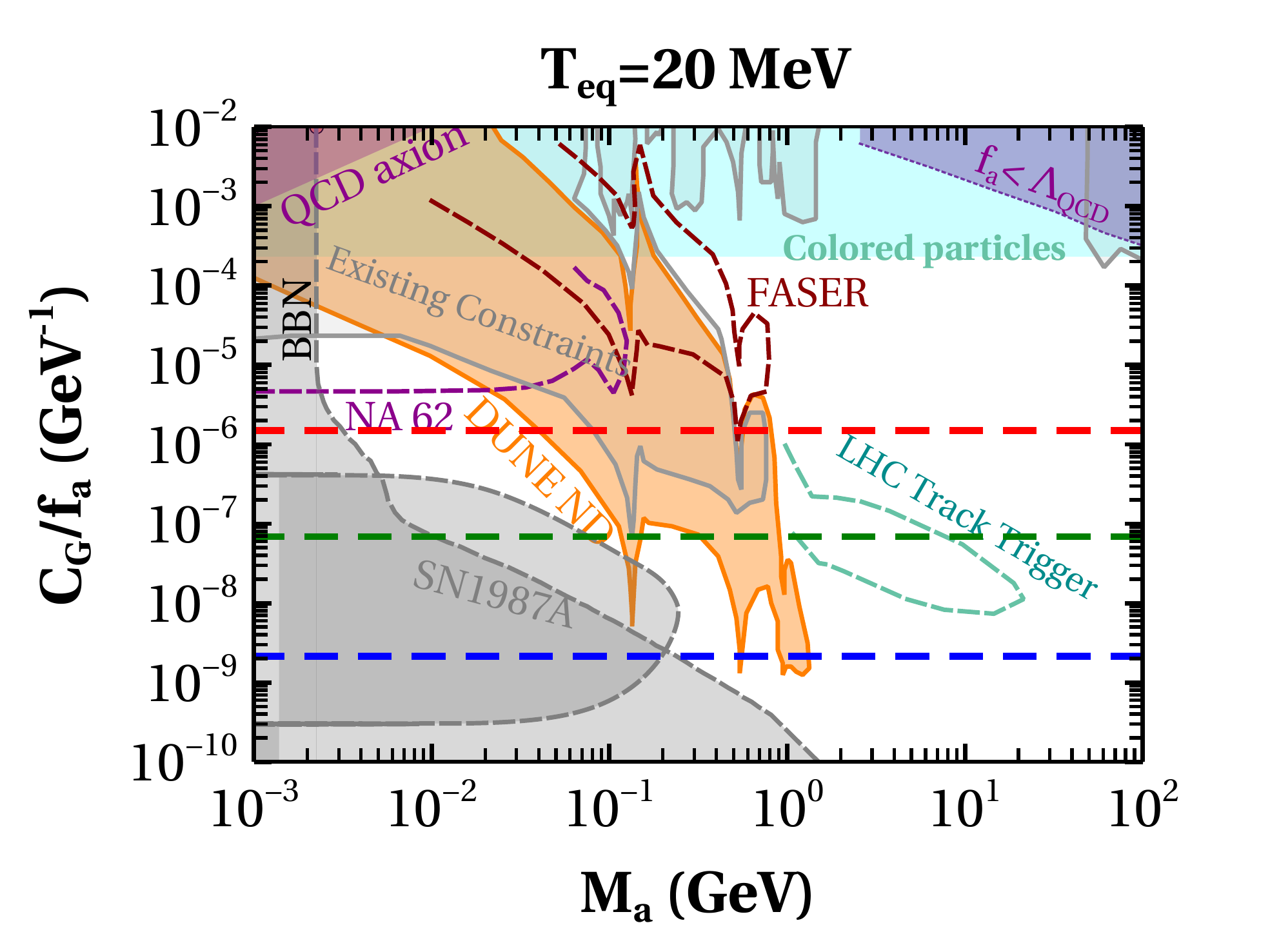}}
    \subfigure[\label{ls_g_2}]{
    \includegraphics[scale=0.52]{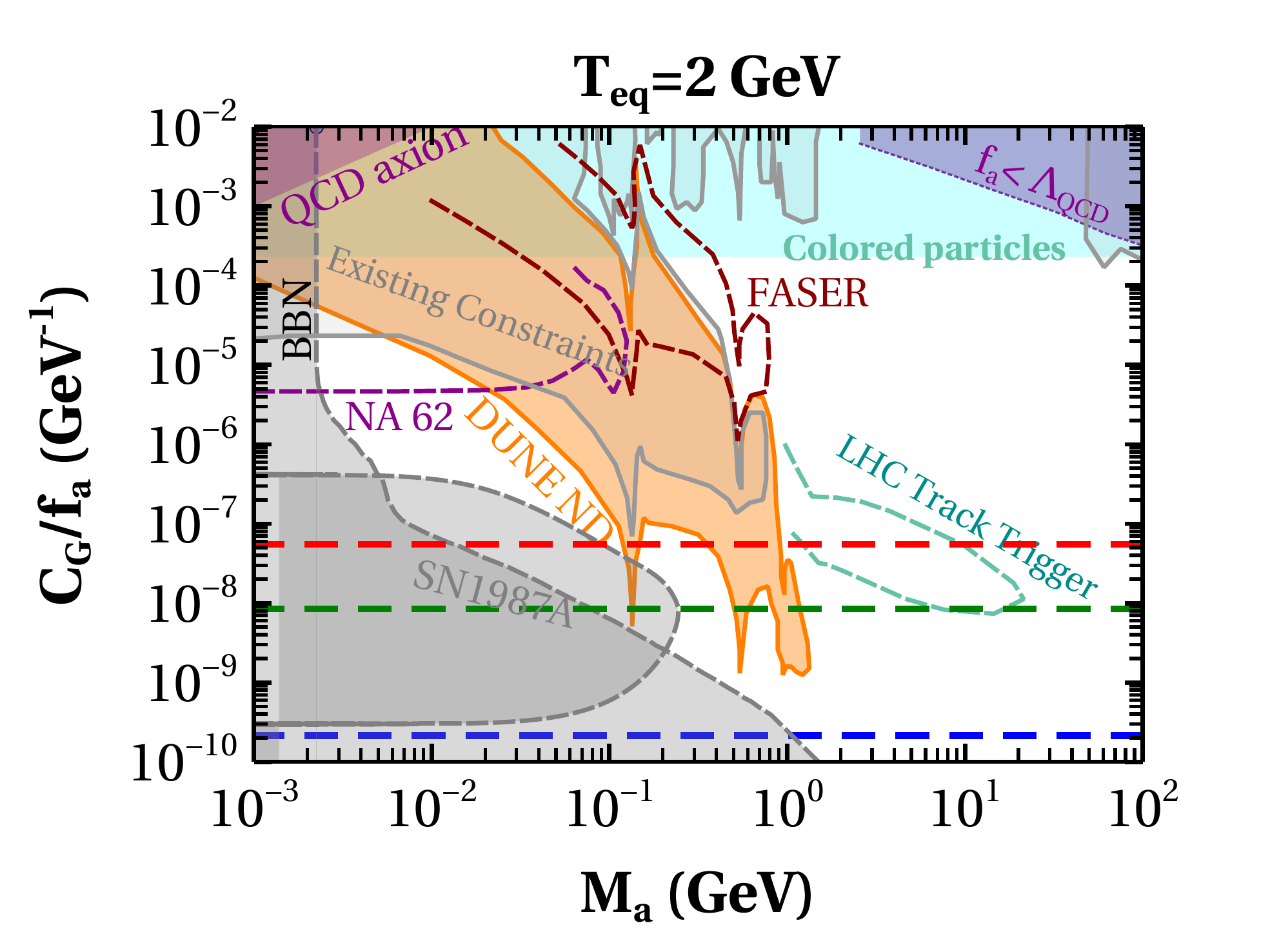}}
    \caption{\it Constraints plot in the $M_a$ versus $C_G/f_a$ plane gluon dominance scenario.
    All kind of astrophysical and collider constraints are shown in different shades. Observed relic satisfying contours are also shown by different horizontal lines in the same plane considering $T_{RH}=10^6$ {\rm GeV} and 
    $C_{\chi}/f_a=10^{-11}$ {\rm GeV}$^{-1}$ for all plots.
    Here we show lines satisfying observed relic density for different values of $n\in\{2,3,4\}$ shown by blue, green and red dashed line respectively for fixed $M_{\chi}=10^3$ {\rm GeV}.
    We considered $T_{eq}=20$ {\rm MeV} in (a) {\bf (top panel)} and 
    $T_{eq}=2$ {\rm GeV} in (b) {\bf (bottom panel)}.
    Relic satisfying lines for $n\leq1$ are beyond existing constraints. For the chosen benchmark point, 
    we obey $(C_G/f_a)_{\rm max} \sim 10^{-5}~\g$ for the EFT consistency of the scenario.}
    \label{fig:lab_search}
\end{figure}

In Fig.\ref{fig:lab_search} we showcase both the constraints and the future experimental reaches in $C_G/f_a - M_a$ 
plane. The horizontal colored lines in the plots are the contours satisfying observed relic density for different values of involved parameter.
For relic satisfying contours we use bench mark values of model parameters, $T_{RH}=10^6$ GeV, $C_{\chi}/f_a=10^{-10}\g$ in both the plots.
We show relic satisfying parameter space with different colored lines for different values of $n$ with $T_{eq}=20$ MeV in 
Fig.\ref{ls_g_1} and $T_{eq}=2$ GeV in Fig.\ref{ls_g_2} respectively for a fixed value of $M_{\chi}=10^3$ GeV.
The contours with different values of $n\in\{2,3,4\}$ are shown by blue, green and red dashed line respectively in the top panel.
From these two plots we notice that the higher the value of $n$
the better the experimental sensitivity as discussed in sec.\ref{sec:relic_ns}.
We also observe that for RD ($n=0$) universe and for $n=1$ (NS) the relic satisfying contours
($C_G/f_a <10^{-10}~\g$)
are far below the existing and future constraints and we do not portray them here.
In our model parameter space satisfying the UV cut-off of the EFT framework, the maximum value of 
$C_G/f_a$ can be $\sim 10^{-5}~\g$.

Comparing Fig.\ref{ls_g_1} and Fig.\ref{ls_g_2} we recognize that for a fixed $n$, the lower the value of $T_{eq}$ the higher the value of $C_G/f_a$ required 
(as discussed in sec.\ref{sec:relic_ns}) and hence better experimental sensitivity. 
From the above figures we make out that though in RD universe the relic satisfying couplings for $M_{\chi}=10^3$ GeV are far below experimental sensitivity but
significant amount of parameter space reproducing observed relic density is excluded in NS cosmology. 
SN1987A and BBN puts most stringent bound for $M_a<100$ MeV and lower values of $C_G/f_a$.
The region “Existing Constraints” include the bounds from partially invisible kaon decays \cite{Ertas:2020xcc}, electron beam dump, CHARM,
visible kaon decays, B decays \cite{Aloni:2018vki}, LHC dijet searches \cite{Mariotti:2017vtv}.
However, there is still some unexcluded parameter space in NS cosmology which can be probed in future experiments like DUNE ND ($C_G/f_a>10^{-9}~\g$), LHC track Trigger ($C_G/f_a>10^{-8}~\g$) and FASER($C_G/f_a>10^{-6}~\g$). In RD universe $M_\chi\lesssim250$ GeV is excluded from EFT consistency as pointed out earlier in this section.

\subsection{Fermion dominance scenario}
In this subsection we consider fermion dominance only where only $C_f/f_a$ and $C_{\chi}/f_a$ non zero while rest of the ALP couplings are zero.
Similar to previous subsection we now present all the constraints and also the parameter space that 
reproduce the observed relic density in $M_a$ - $C_f/f_a$ plane.

\begin{figure}[tbh!]
    \centering
    \subfigure[\label{ls_f_1}]{
    \includegraphics[scale=0.56]{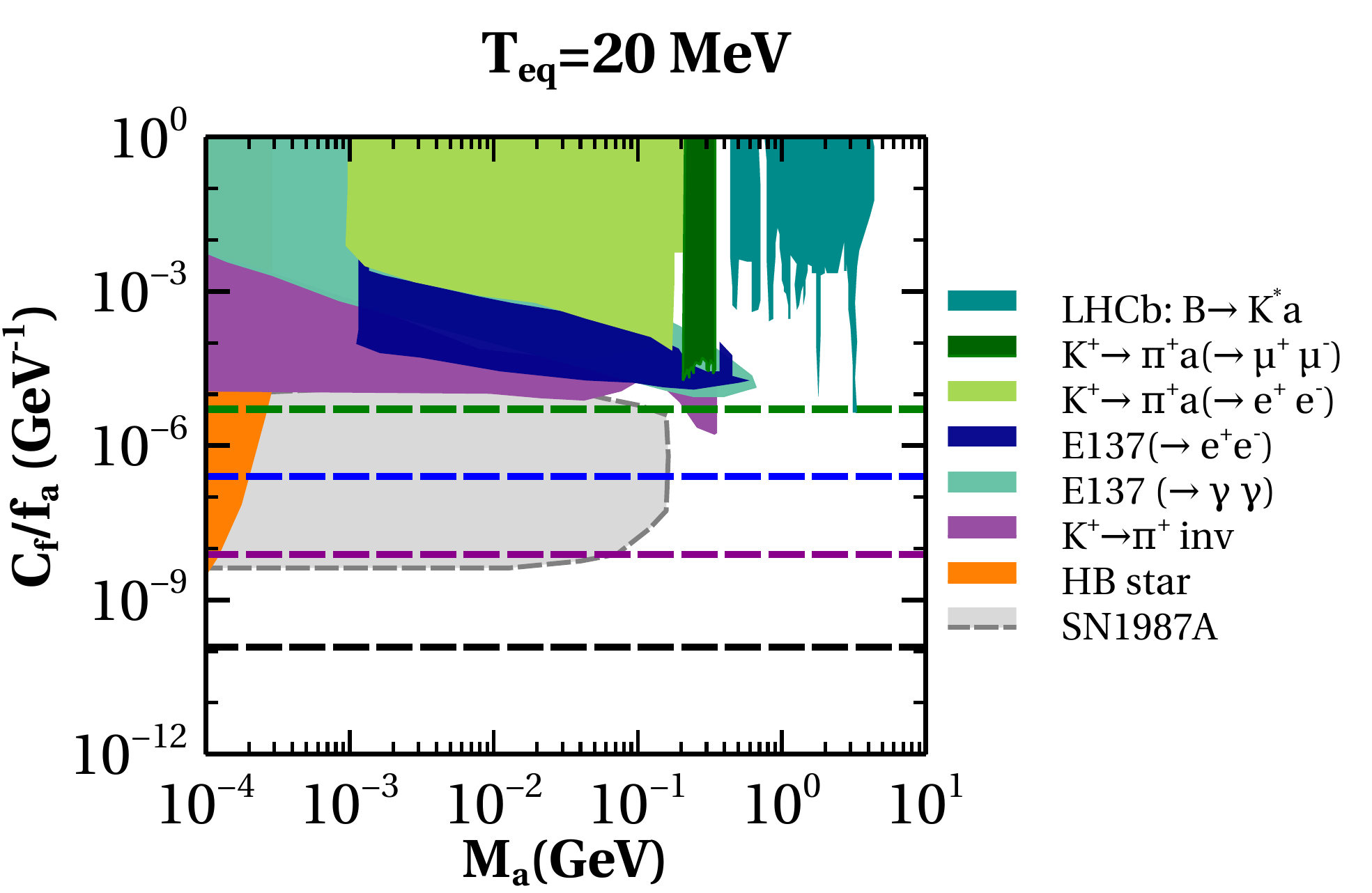}
   }
    \subfigure[\label{ls_f_2}]{
    \includegraphics[scale=0.56]{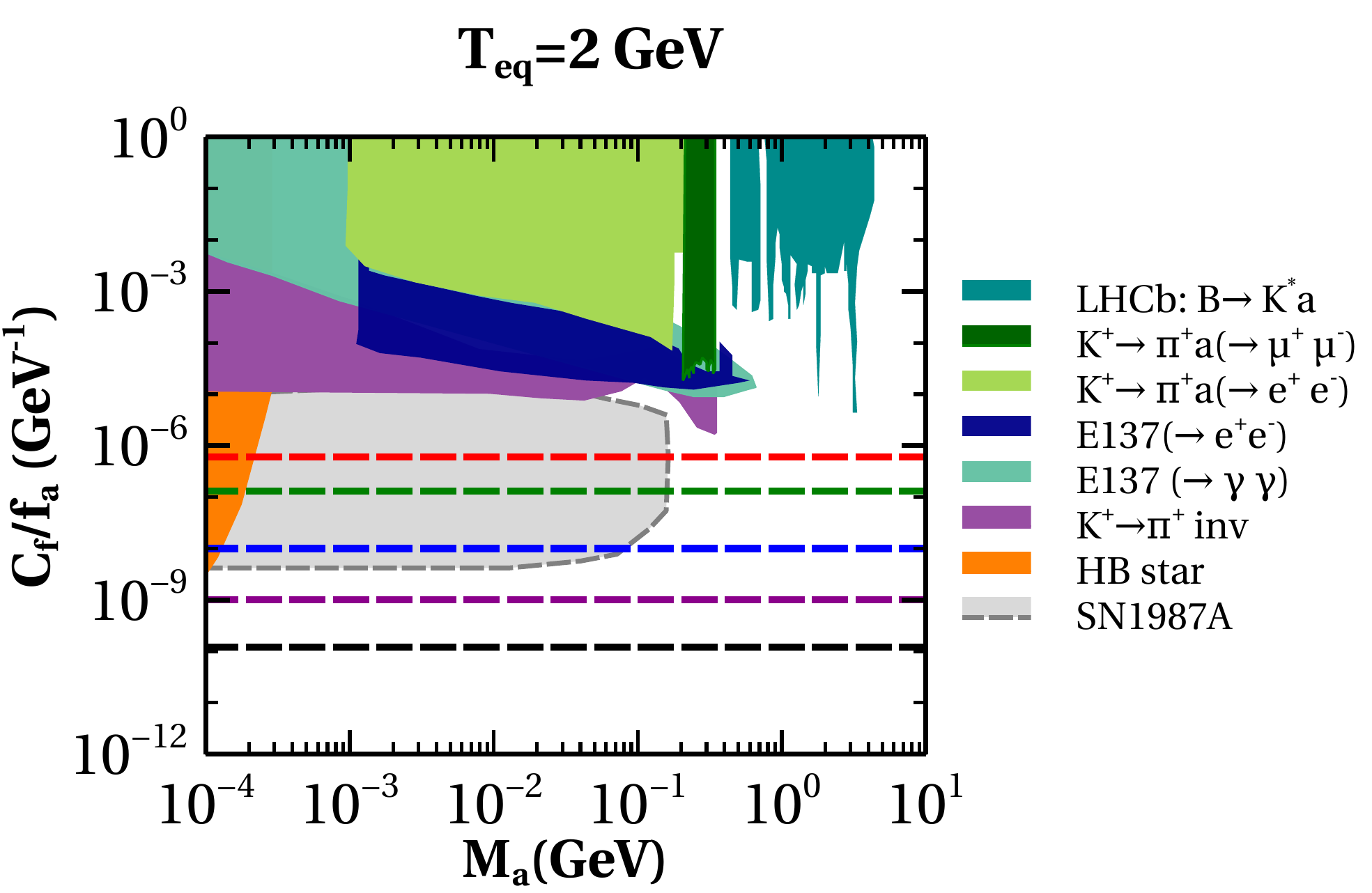}
    }
    \caption{\it Constraints plot in the $M_a$ versus $C_f/f_a$ plane.
    Existing astrophysical and collider constraints are shown in different shades. Observed relic satisfying contours are also shown by different horizontal lines in the same plane considering $T_{RH}=10^6$~{\rm GeV}, 
    $C_{\chi}/f_a=10^{-11}$~{\rm GeV}$^{-1}$, $M_{\chi}=10^3$~{\rm GeV} with  (a) {\bf (top panel)} $T_{eq}=20$ {\rm MeV} and (b) {\bf (bottom panel)}  $T_{eq}=2$~{\rm GeV}.
    We show lines satisfying observed relic density for different values of $n\in\{0,1,2,3,4\}$ shown by black,magenta blue, green and red dashed line respectively. 
    For the chosen benchmark point, 
    $(C_f/f_a)_{\rm max} \sim 10^{-5}~\g$ is set for the EFT consistency of the scenario.
    In (a) relic satisfying line corresponding $n=4$ is above the upper limit of
    $C_f/f_a$ and is not portrayed here.
    }
    \label{fig:f_lab_search}
\end{figure}

Before going to the discussion of future detection prospects we first elaborate on the various existing constraints relevant on the the ALP parameter space ($M_a - C_f/f_a$) in Fig.\ref{fig:f_lab_search}.
The horizontal colored lines in  Fig.\ref{ls_f_1} \& Fig.\ref{ls_f_2} are the contours satisfying the 
observed relic density with fixed values of $T_{RH}=10^6$ GeV
and $C_{\chi}/f_a=10^{-10}~\g$ respectively.
We display the relic density contours with different colored lines for different values of $n$ with $T_{eq}=20$ MeV  and $T_{eq}=2$ GeV in 
Fig.\ref{ls_f_1} \& Fig.\ref{ls_f_2}
 respectively with fixed value of $M_{\chi}=10^3$ GeV.
The contours with different values of $n\in\{0,1,2,3,4\}$ are shown by black, magenta, blue, green and red dashed line respectively.
From these two plots  we observe that higher values of $n$
give better experimental sensitivity keeping other parameters fixed, as discussed in sec.\ref{sec:relic_ns}.
Here also we claim that for RD ($n=0$) universe the relic satisfying contours
($C_f/f_a <10^{-10}~\g$)
are far below the existing constraints as expected from the thermalization
condition discussed in sec-\ref{sec:relic}.
The observed relic density satisfying line corresponding to $n=4$ in Fig.\ref{ls_f_1} is above the upper limit of $C_f/f_a$ and we do not portray that line keeping the EFT analysis valid.
From this plot we conclude that for a fixed $n$ lower values of $T_{eq}$
provide better experimental sensitivity than higher values of $T_{eq}$
(as discussed in sec.\ref{sec:relic_ns}). 

Just like gluon dominance, the most stringent bound in this case also comes from SN1987A which excludes $C_f/f_a$ upto $10^{-9}~\g$ for $M_a<100$ MeV.
The limits obtained from various collider experiments like E137,invisible decay of Kaon and LHCb stifle higher values of $C_f/f_a$ for $M_a$ ranging from KeV to few GeV.
From Fig.\ref{fig:f_lab_search} we again surmise that the relic satisfying couplings for higher DM mass ($M_{\chi}\sim1$TeV)  are far below experimental sensitivity in RD universe. 
However, sizable portion of contours reproducing observed relic density is excluded in NS cosmology.

\begin{figure}[tbh]
    \centering
    \subfigure[\label{ls_fp_1}]{
    \includegraphics[scale=0.5]{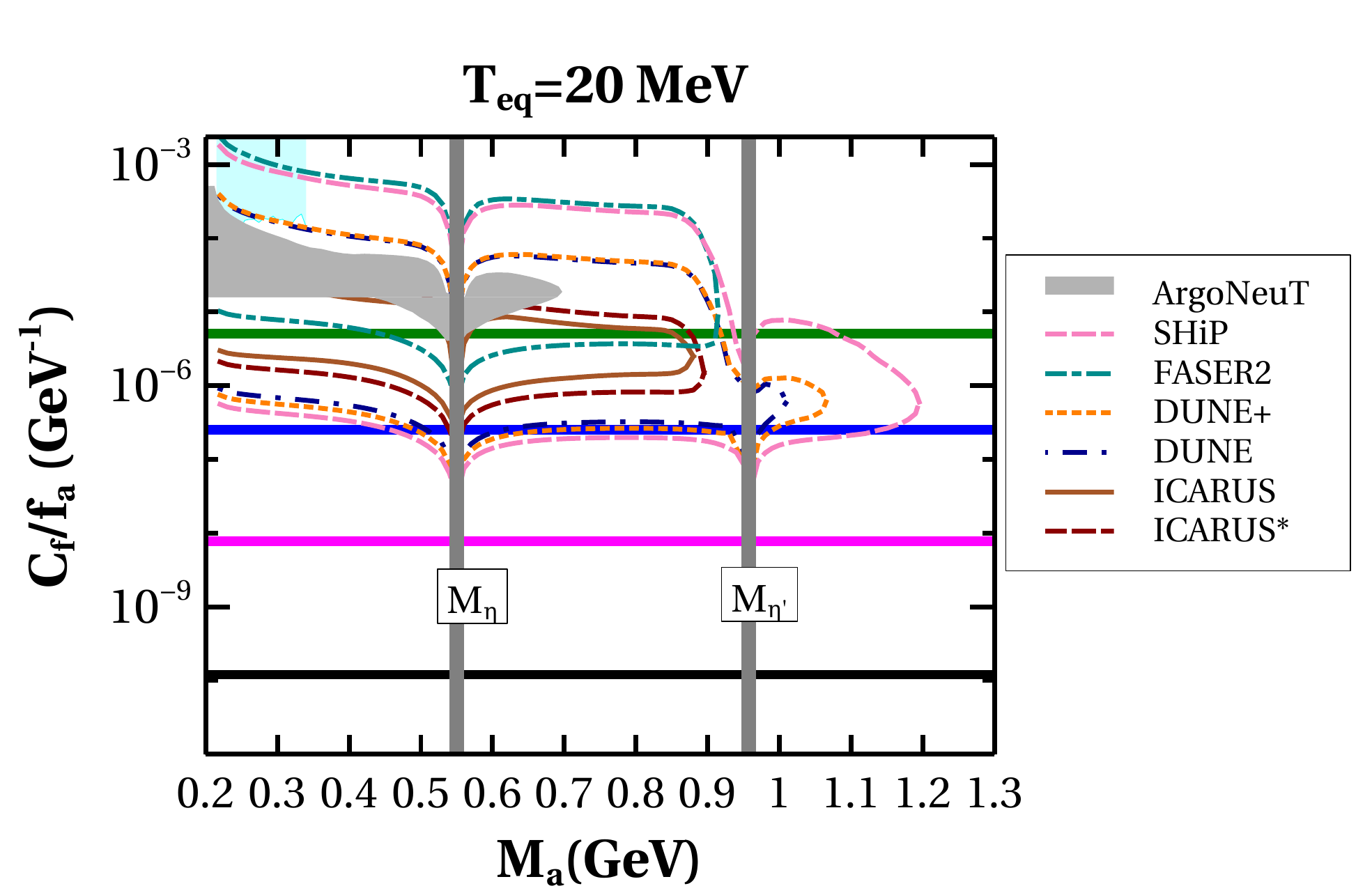}
    }
    \subfigure[\label{ls_fp_2}]{
    \includegraphics[scale=0.5]{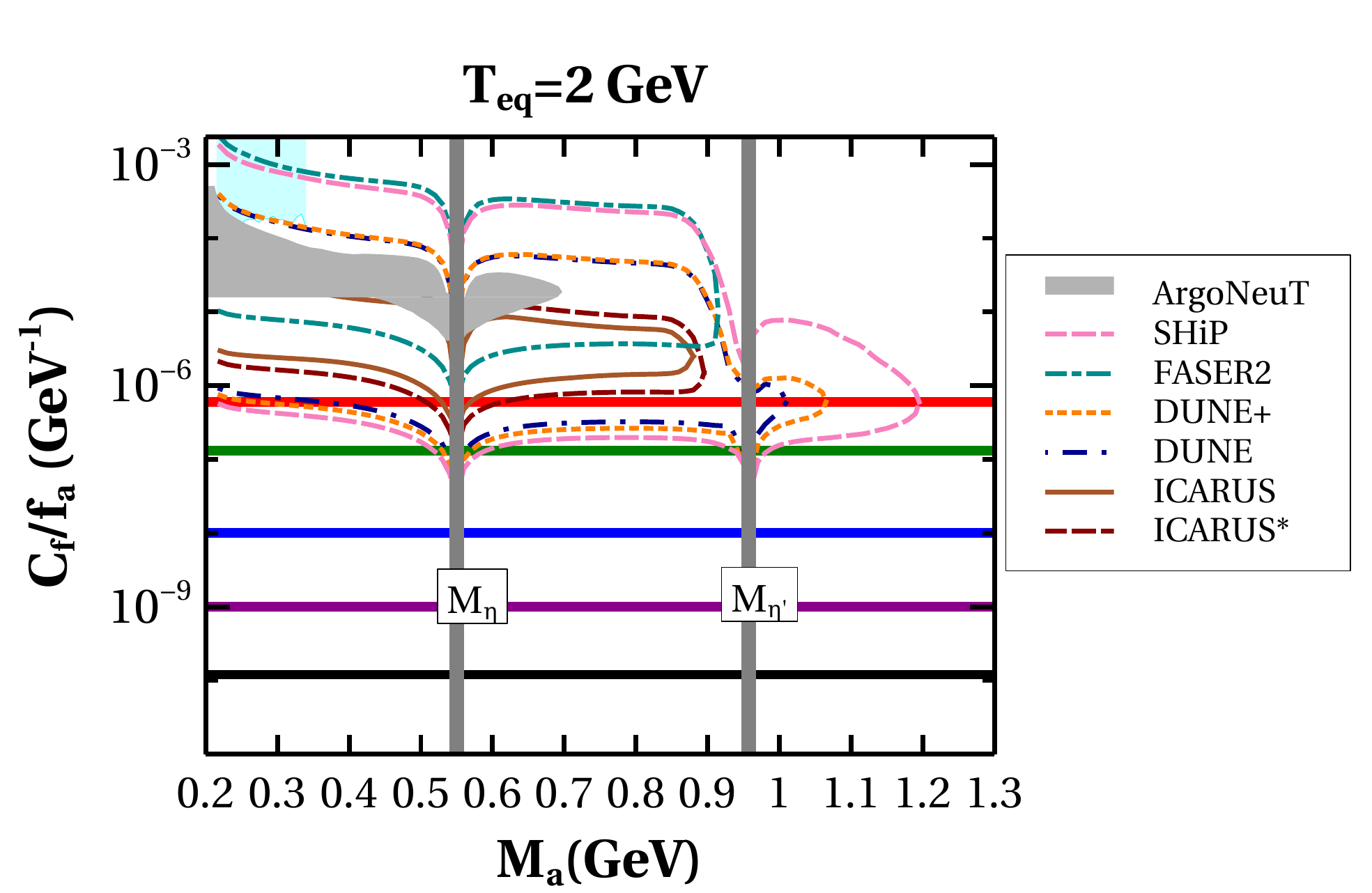}
    }
    \caption{\it Future probes from laboratory searches in the $M_a$ versus $C_f/f_a$ plane for
     $T_{RH}=10^6$ {\rm GeV}, $C_{\chi}/f_a=10^{-11}$ {\rm GeV}$^{-1}$, $M_{\chi}=10^3$ {\rm GeV} with (a) $T_{\rm eq}=20$ {\rm MeV} and (b) $T_{\rm eq}=2$ {\rm GeV}. 
    The thick solid colored lines depict relic satisfying contours for $n\in(0,1,2,3,4)$ in black, magenta, blue, green, red color respectively. 
    For the chosen benchmark point, we set
    $(C_f/f_a)_{\rm max} \sim 10^{-5}~\g$  for the EFT consistency of the scenario.
    In (a) relic satisfying line corresponding $n=4$ is above the upper limit of
    $C_f/f_a$ and is not portrayed  here.}
    \label{fig:f_lab_search_3}
\end{figure}

Inspite of all of these stringent limits, there is still some unexcluded parameter space providing observed relic density in NS cosmology which 
can be probed in future generation experiments.
In Fig.\ref{fig:f_lab_search_3} we present such future projection along with relic satisfying contours in 
$M_a$ vs. $C_f/f_a$ plane. There is a mass window for $M_a=0.2-1.3$ GeV 
which can be explored in future as discussed in ref.\cite{Co:2022bqq} and we present the same in the context of our analysis. 
Here, also in Fig.\ref{fig:f_lab_search_3} the horizontal colored lines in all plots depict contours satisfying the 
observed relic density with fixed values of $T_{RH}=10^6$ GeV
, $C_{\chi}/f_a=10^{-10} ~\g$ and $M_{\chi}=10^3$ GeV respectively.
We exhibit the relic density contours with different colored lines for different values of $n$ with $T_{eq}=20$ MeV in 
Fig.\ref{ls_fp_1} and $T_{eq}=2$ GeV in Fig.\ref{ls_fp_2}
respectively with fixed value of $M_{\chi}=10^3$ GeV.
The black, magenta, blue, green and red solid lines signify different values of $n\in\{0,1,2,3,4\}$  respectively.
In Fig.\ref{fig:f_lab_search_3} we just focus on the mass range $M_a\sim0.2-1.3$ GeV and all the dependence of involved parameters on relic density is same as in Fig.\ref{fig:f_lab_search}.
In Fig.\ref{ls_fp_1} also, the observed relic density satisfying line corresponding to $n=4$  is above the upper limit of $C_f/f_a$ and we do not portray  that line for the consistency of the EFT analysis.
For brevity we don't discuss again the features of the contours satisfying relic density here.
However, it is easy to infer that, while holding all other parameters constant, contours characterized by higher values of $n$ and lower values of $T_{eq}$ are more sensitive to future projections.



The gray shaded region in all the plots in Fig.\ref{fig:f_lab_search_3} are excluded from the 
ArgoNeuT experiment at Fermilab where ALP is 
produced from neutrino beam target and the subsequent search from its decay into di-muon pairs \cite{ArgoNeuT:2022mrm}. 
Experiments like  SBND\cite{MicroBooNE:2015bmn}, ICARUS\cite{MicroBooNE:2015bmn}, 
SHiP \cite{SHiP:2015vad} and  FASER 2 \cite{Feng:2017uoz} will be able to probe a significant amount of parameter space reproducing observed relic density in NS cosmology.
The label ``ICARUS$^*$'' signify ICARUS experiment with off axis 
neutrino beam  at the Main Injector \cite{Co:2022bqq}.
The label ``DUNE$+$'' depicts the result of DUNE experiment
with additional $30$m length before the front detector panel as part of the decay volume as discussed in ref.\cite{Co:2022bqq}. For the chosen benchmark parameters $M_\chi\lesssim 480$ GeV is excluded from EFT consistency for fermion dominance.

Therefore, we demonstrate that our proposed frame work offers greater verifiability in future generation experiments 
\footnote{
See Refs. \cite{Dolan:2017osp,Kelly:2020dda,Bharucha:2022lty,Co:2022bqq} 
for more discussions on existing constraints and future projections for ALP.}
for both gluon dominance and fermion dominance scenarios when the fast expansion parameter $n$ takes higher values. This was the primary motivation behind this work.

\section{Direct Detection}
\label{sec:dd}
The DM direct detection signal can be obtained in our set up through DM nucleon elastic scattering and this
elastic scattering amplitude can be calculated using the following effective Lagrangian 
\begin{equation}
\mathcal{L}_{eff}= \left( \frac{C_G}{f_a}\right) \left( \frac{C_{\chi}}{f_a}\right)\frac{1}{M_a^2} M_{\chi} \Bar{\chi}\gamma_5 \chi \,G_{a\mu \nu}\Tilde{G}^{a\mu \nu},  
\end{equation}
where, we assume the momentum transfer $q <<M_a$. The presence of explicit $\gamma_5$ in the
effective interaction vertex will lead to momentum $(q)$ suppression in non-relativistic limit in 
the scattering amplitude\cite{Hisano:2015bma}. Moreover, the $a\,G\,\tilde {G}$ vertex will also 
reduce to a current, containing $\gamma_5$ which will also lead to a another momentum $q$ suppression
\cite{Cheng:2012qr,Shifman:1978zn}. Thus the overall cross-section will contain
a suppression of $q^4$, a characteristic feature of direct detection mediated by pseudo-scalar
\cite{Abe:2018emu,Arcadi:2017wqi,Banerjee:2017wxi}. 
Apart from the small effective couplings ($C_G/f_a, C_{\chi}/f_a$) required to produce $\chi$ non thermally the direct detection cross-section gets a suppression of $q^4$ where as typical value of $q^2$ in current experiments are 
$\mathcal{O}(10 \text{KeV})$ \cite{XENON:2018voc} and hence evade the current experimental sensitivity.

\medskip
\section{Discussion and Conclusions}
\label{sec:concl}

In this work we show that the ALPs can act as the portal to the dark sector, particularly  we emphasize that the present and upcoming ALP-search experiments may serve as tools for investigating ALP-portal freeze-in dark matter.
Working in an effective theory framework in both radiation-dominated early universe and other non-standard cosmological histories we study the parameter space satisfying the observed DM relic density.
We identify detectable regions in such parameter space to be within the reach of future ALP search experiments. From our investigations we summarize our major findings below: 
\begin{itemize}
    
    \item DM relic density is proportional to the effective couplings $C_G/f_a$,$C_f/f_a$,$C_{\chi}/f_a$ and $M_{\chi}$ (mass of DM) as shown in eq.\eqref{eq:ydm_g} and eq.\eqref{eq:ydm_f}. In our analysis we consider higher-dimensional non-renormalizable operators between dark sector and visible sector via the ALP-portal, therefore the DM production has dependence on $T_{RH}$ (see eq.\eqref{eq:ydm_g}, \eqref{eq:ydm_f}). We find that the relic density is also proportional to $T_{RH}$ (see Fig.\ref{fig:dm}).
    However, for our chosen parameter space the relic density is almost insensitive to the light mediator $M_a$ due to freeze in happening at high $T_{RH}$. 
    Conditions from the non-thermalization of $\chi$ sector with the SM bath imposes a strict limit from the DM production on the effective couplings as decided by $M_{\chi}$ dependencies in eq.\eqref{eq:therm1} and eq.\eqref{eq:therm2}. Our analysis shows that for $M_{\chi}\lesssim 250$ GeV ($480$ GeV) the effective couplings are already excluded (from EFT consistency) for gluon (fermion) dominance, whereas $M_{\chi}\sim 1$ TeV the effective couplings are too small to be probed by the present and future generation experiments. 
    
    \item 
    The aforementioned situation changes drastically if we assume a non standard species $\phi$ contributing significantly to the energy density of early universe at an early epoch. 
    In such scenario the expansion rate of the universe $H$ is modified which leads to a faster expansion of the universe than the radiation-dominated universe.
    In such a framework the value of $H$ allows larger values of effective couplings ($C_G/f_a$,$C_f/f_a$,$C_{\chi}/f_a$) still keeping the dark sector non thermalized. The relaxation of parameter space can be understood by comparing
    Fig.\ref{fig:th} and Fig.\ref{fig:therm_ns}
   where we show that the effective couplings can be $\mathcal{O}(10^2)$  larger in NS cosmological scenario than in RD universe. In the two aforementioned figures we also show the parameter space which leads to {\it sequential freeze-in} and present the allowed parameter space from EFT consistency.
    After solving respective Boltzmann equations with modified Hubble $H$ 
    we are able to identify the parameter space satisfying the observed relic density.
    This boosts the detection 
    prospects for the scenario in laboratories. 

    \item The non-standard cosmology allows the involved effective SM-ALP couplings to be large in order to satisfy the observed relic, therefore this can be probed in present and future generation experiments. 
    Moreover from Fig.\ref{fig:lab_search} and Fig.\ref{fig:f_lab_search} we find that higher the value of n, the better the detection prospects for ALP-SM couplings. 
    The most stringent bound on the relic satisfying parameter comes from astrophysical probes specially SN1987A and HB stars.
    The value of $\Delta N_{eff}$ (dark radiation) estimated from BBN gives a lower bound on the ALP mass $M_a$. Inspite of taking into consideration the existing 
    bounds on $C_G/f_a$ and $C_f/f_a$ coming from electron beam dump, CHARM, visible kaon decays, B decays and LHCb, etc. 
    There remains still leaves some unexcluded parameter space satisfying the observed DM relic density. 
    The DUNE and LHC track trigger will be able to probe such experimental parameter space that correspond to $C_G/f_a\sim 10^{-5}-10^{-9}~\g$ for the mass range of $M_a\sim 0.1-10$ GeV as shown in Fig.\ref{fig:lab_search}. 
    Similarly, experiments like SHiP, DUNE, FASER2, ICARUS will be able to probe $C_f/f_a\sim 10^{-4}-10^{-6}~\g$ for the mass range of $M_a\sim 0.2-1$ GeV as displayed in Fig.\ref{fig:f_lab_search_3}.
    
    \item Finally we point out that our analysis is also an alternative probe for non-standard cosmological evolution of early universe. In Fig.\ref{fig:n_Tr}
    we show that for a fixed ($C_G/f_a$), ($C_f/f_a$) and $C_{\chi}/f_a$ one is probing values of $n$ and $T_{eq}$ which satisfy the observed DM relic density. The detection prospects of various effective couplings for our proposed 
    scenario are also highly sensitive to $n$ and $T_{eq}$.
    For example, comparing Fig.\ref{ls_g_1} and \ref{ls_g_2} we note that the future probable regions of relic satisfying contours change with $T_{eq}$ for the case of gluon dominance. Similar conclusion can be drawn for the fermion dominance case by 
    comparing Fig.\ref{ls_fp_1} and \ref{ls_fp_2}.
    Hence any signal in the upcoming experiments will provide us concrete information regarding the cosmic evolution and dark sector assuming ALP-portal freeze-in DM scenario. 
\end{itemize}

Although we have several possible ways by which the parameter space for ALP-mediated FIMPs can be extended, some of these are already excluded by the current experimental limits. 
For example, heavy ALPs which mediate the entropy transfer through off-shell interactions (the interactions we considered \& other interactions including CP violating ALP interactions) are excluded by the LEP constraints.
One can explore scenarios where ALPs do not couple to all fermions through a universal Yukawa coupling; rather ALPs couple only to electrons (electrophilic) or to charged leptons. In such case the constraints are  from SLAC 137 and the muon anomalous magnetic moment is less stringent which is not addressed so far \cite{Essig:2010gu,Bauer:2017ris}. 
 Because of the great interest in the ALP search communities we envisage in near future more dedicated search facilities will be launched to probe ALP parameter space and thus also paving the way for detectable ALP-portal freeze in DM as well.

\section*{Acknowledgement}

AG and SJ thank Soubhik Kumar for providing data for collider and astrophysical bounds on ALP couplings. 
DKG and SJ would like to thank Aoife Bharucha for the enlightening discussion. 
SJ thanks Sougata Ganguly and Tanmay Kumar for valuable discussions.
SJ acknowledges financial support from CSIR, Government of India, under the NET JRF fellowship scheme
with Award file No. 09/080(1172)/2020-EMR-I.
\appendix
\section{UV freeze-in from Gluon fiusion}
\label{apxc}
For the case of Gluon dominance scenario, freeze in production of DM from SM bath takes place via the following two operators :  
\begin{equation}
\mathcal{L_{\rm ALP-SM}}\supset - \frac{C_{G}}{f_a}a\,G_{a\mu \nu}\Tilde{G}^{a\mu \nu}     
\end{equation}
\begin{equation}
\mathcal{L}_{ALP-\chi}=-\frac{C_{\chi}}{f_a} \, (\Bar{\chi}\gamma^{\mu}\gamma^5 \chi) \partial_{\mu}a    
\end{equation}
Including the color factor the amplitude square will be,  
\begin{equation}
|M|^2_{\rm gg-\chi\chi}= 256 \dfrac{C_{G}^2 C_{\chi}^2 M_{\chi}^2 s^3}{f_a^4 (s-M_a^2)^2} ,   
\end{equation}
where, $s$ is the center of mass energy.
The cross-section will be
\begin{equation}
\sigma_{gg\rightarrow \chi \chi}= \dfrac{16 C_{\chi}^2 C_{G}^2 M^2_{\chi} s^2}{\pi f_a^4 (M_a^2 -s)^2} \sqrt{1- \dfrac{4 M_{\chi}^2}{s}}
\label{eq:sig1}
\end{equation}
Consider the process: $g+g\rightarrow \chi +\chi$
The Boltzmann equation is given by,
\begin{equation}
  \dot{n}_{\chi} + 3 H n_{\chi} = \int d\Pi_g d\Pi_g d\Pi_{\chi} d\Pi_{\chi} \delta^4(p_1+ p_2 -p_3-p_4) |M|^2_{\rm gg\to\chi\chi} \mathcal{F}_g(p_1)~ \mathcal{F}_g(p_2) ,
  \label{eq:cf}
\end{equation}
where, $p_{1,2}$ are the momentum of incoming gluons and $p_{3,4}$ are the momentum of outgoing $\chi$ particles. $\mathcal{F}_g(p_i)$ denotes the distribution function of gluons. 

\subsection{Using MB distribution:}

We assume initial bath particles follow the MB distribution, $\mathcal{F}_g \sim e^{-E_g/T}$. The term in the R.H.S of eq \eqref{eq:cf} is the collision term $C[\mathcal{F}_g]$.
After some simplification the collision term looks like \cite{Elahi:2014fsa}:
\begin{eqnarray}
C[\mathcal{F}_g]&=& \frac{ T}{512 \pi^6}\, 4\pi\, \int_{4 M_{\chi}^2}^{\infty} ds \mathcal{P}_{gg}  \mathcal{P}_{\chi \chi}\, |M|^2 \, \frac{1}{\sqrt s} K_{1}\left( \frac{\sqrt s}{T} \right)
\label{eq:cf2}
\end{eqnarray}
where, $\mathcal{P}_{ij} $ is given by,
\begin{equation}
\mathcal{P}_{ij}= \frac{1}{2 \sqrt{s}} \sqrt{s-(m_i+m_j)^2} \sqrt{s-(m_i-m_j)^2}
\end{equation}

Upon proper substitution in eq. \eqref{eq:cf2} we get,

\begin{eqnarray}
\nonumber
C[\mathcal{F}_g]& \approx &  \frac{  T}{128 \pi^5} \, \frac{256}{f_a^4} \,C_{G}^2 C_{\chi}^2 M_{\chi}^2\,\int_{4 M_{\chi}^2}^{\infty} ds \frac{1}{4} \sqrt{(s- 4 M_{\chi}^2)s}\, \dfrac{s^3}{(s-M_a^2)^2} \, \frac{1}{\sqrt s} K_{1}\left( \frac{\sqrt s}{T} \right)\\
&\approx & \frac{ T^6}{2 \pi^5}\dfrac{32 C_{G}^2 C_{\chi}^2 M_{\chi}^2}{f_a^4},
\label{eq:cf3}
\end{eqnarray}

\subsection{Using BE distribution}
Using the proper quantum correction the collision term takes the following form \cite{Lebedev:2019ton},
\begin{eqnarray}
\nonumber
C_{rel}[\mathcal{F}_g]&=&\int d\Pi_g d\Pi_g d\Pi_{\chi} d\Pi_{\chi} \delta^4(p_1+ p_2 -p_3-p_4) |M|^2_{\rm gg\to\chi\chi}\\
&&\times \mathcal{F}_g(p_1) \mathcal{F}_g (p_2)~ ,
\label{fd1}
\end{eqnarray}
where,  $\mathcal{F}_g$ is now BE distribution function and we write it as \cite{Lebedev:2019ton},
\begin{equation}
 \mathcal{F}_g (p)=\dfrac{1}{ e^{\frac{V.p}{T}}\pm 1} ,  
\end{equation}
where, $V_{\mu}$ is the 4-velocity in the center of mass frame 
The cross section for the process $gg\to \chi \chi$ is given by,
\begin{equation}
\sigma_{gg\to \chi \chi} = \frac{1}{4~E_1 E_2 v_{mol}} \int d\Pi_{\chi} d\Pi_{\chi} \delta^4(p_1+ p_2 -p_3-p_4) |M|^2_{\rm gg\to\chi\chi}, 
\label{fd_sig}
\end{equation}
where $E_{1,2}$ initial sate particle's energy and $v_{mol}$ is the Moller velocity.
$v_{mol}$ is given by,
\begin{equation}
v_{mol} =\frac{p_1.p_2}{E_1~E_2}
\end{equation}
So, incorporating $v_{mol}$ in eq.\eqref{fd_sig} one obtains the following equality:
\begin{equation}
\sigma_{gg\to \chi \chi} v_{mol} = \frac{1}{4~p_1 p_2 } \int d\Pi_{\chi} d\Pi_{\chi} \delta^4(p_1+ p_2 -p_3-p_4) |M|^2_{\rm gg\to\chi\chi}, 
\label{fd_sig1}
\end{equation}
Now we define the collision term as 
\begin{eqnarray}
\nonumber
C_{rel}[\mathcal{F}_g]&=&\int d\Pi_1 ~d\Pi_2 4 p_1 p_2~\sigma_{gg\to \chi \chi} v_{mol} ~\mathcal{F}_g(p_1) \mathcal{F}_g (p_2)~ 
\label{fd2}
\end{eqnarray}

Following the notation in ref.\cite{Lebedev:2019ton}, in gas rest frame we define
\begin{equation}
    p=\frac{p_1+p_2}{2},~ k=\frac{p_1-p_2}{2},
\end{equation}
where, $p$ is related to c.o.m frame by a Lorentz transformation $p= \Lambda (E,0,0,0)$ , $E$ being the c.o.m energy and $\Lambda$ is the Lorentz transformation.
$\Lambda$ is function of rapidity $\eta$  and angular variables $\theta,\phi$ as given in \cite{Lebedev:2019ton}. 
Similarly $k$ can also be translated to c.o.m frame 
$k= \Lambda(\eta,\theta_k,\phi_k) (0,l_1,l_2,l_3)$, where $|l_i|=|\Vec{p_1}-\Vec{p_2}|$.
where $\theta_k,\phi_k$ are angular variable associated with $k$-space.
So, for mass less particle $|l_i|=E$. For convenience we consider the momentum along only one direction as $k= \Lambda(\eta,\theta_k,\phi_k) (0,E,0,0)$.

Following the formalism as in ref.\cite{Lebedev:2019ton}
we can write the distribution functions as
\begin{eqnarray}
 \mathcal{F}_g (p_1)&=&\dfrac{1}{ \exp{[\frac{1}{T}(E \cosh \eta + E \sinh \eta \cos \theta_k)]}- 1}, \\
 \mathcal{F}_g (p_2)&=&\dfrac{1}{ \exp{[\frac{1}{T}(E \cosh \eta - E \sinh \eta \cos \theta_k)]}- 1}.
\end{eqnarray}
After little algebra we can also write the collision term in eq.\eqref{fd2} accordingly \cite{Lebedev:2019ton},
\begin{eqnarray}
 \nonumber
C_{rel}[\mathcal{F}_g]&=&\int_{M_\chi}^{\infty} dE~  E^3 ~ 4 ~p_1 ~p_2 ~\sigma_{gg\to \chi \chi} v_{mol}\int_{0}^{\infty} d\eta ~\sinh^2 \eta~ \int ~d\Omega_p ~d\Omega _k \\
&&\times \mathcal{F}_g(p_1) \mathcal{F}_g (p_2)~, 
\label{eq:fdf}
\end{eqnarray}
where $d\Omega_i\equiv \sin\theta_i~ d\theta_i~ d\phi_i$ is the solid angle.
We perform the solid angle integration first and get, 
\begin{eqnarray}
 \nonumber
 \int~d\Omega_p ~d\Omega _k  \mathcal{F}_g(p_1) \mathcal{F}_g (p_2)~ &=&
 \dfrac{1}{E~\sinh \eta ~(\exp{[\frac{1}{T}(E \cosh \eta)]}-1)}\\
 &&\times \ln \left( \dfrac{ \frac{\sinh \eta}{2T}(E \cosh \eta +E \sinh \eta) }{\frac{\sinh \eta}{2T}(E \cosh \eta -E \sinh \eta)} \right)
\end{eqnarray}
Incorporating this in eq. \eqref{eq:fdf} we will find $\Gamma_{gg\to \chi \chi}=1/n^{eq}_g~C^{rel}[\mathcal{F}_g]$ using BE distribution.
\begin{figure}[tbh!]
    \centering
  
    \includegraphics[scale=0.5]{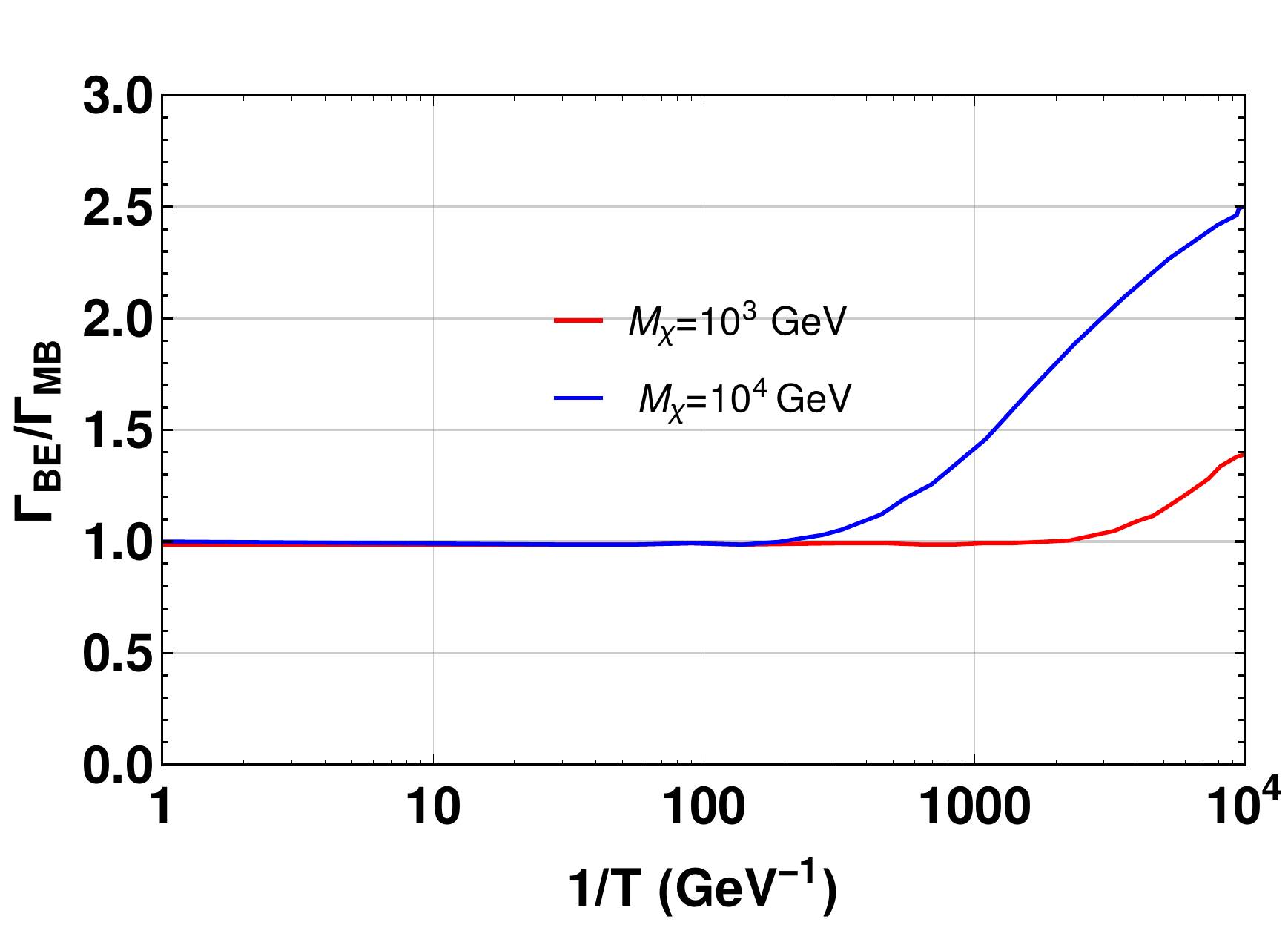}
    \caption{Evaluation of the ratio $\Gamma_{\rm BE}/\Gamma_{\rm MB}$ with $1/T$ (GeV$^{-1}$) for $C_{\chi}/f_a=10^{-12}\g,~C_G/f_a=10^{-10}\g$ using two different values of $M_{\chi}=10^3,~10^4$ GeV shown by the red and blue solid lines.}
    \label{fig:mb_be}
\end{figure}

In Fig.\ref{fig:mb_be} we compare the two interaction rates for MB distribution and BE distribution. We denote $\Gamma_{\rm BE}$ and $\Gamma_{\rm MB}$ as interaction rates for $g+g\to \chi+\chi$ using BE and MB distribution function respectively. 
We show the evolution of the ratio $\Gamma_{\rm BE}/\Gamma_{\rm MB}$ with $1/T$ (GeV$^{-1}$) for $C_{\chi}/f_a=10^{-12}\g,~C_G/f_a=10^{-10}\g$ using two different values of $M_{\chi}=10^3,~10^4$ GeV depicted by the red and blue solid lines.
We notice that at higher temperature only, the ratio differs from $1$ i.e. the two interaction rates $\Gamma_{\rm BE}$ and $\Gamma_{\rm MB}$ differ by a factor of $\sim 3$.
However, at low temperature the ratio becomes $1$.
This justifies our  approximation  using MB distribution.
\subsection{Co-moving abundances}
From now on we  use MB distribution to compute the collision term. 
So, we substitute the collision term from eq.\eqref{eq:cf3} in eq.\eqref{eq:cf} and
the Boltzmann equation becomes,
\begin{eqnarray}
\frac{d Y_{\chi}}{d T} (-sHT)&=&  \frac{16 T^6}{ \pi^5}\dfrac{C_{G}^2 C_{\chi}^2 M_{\chi}^2}{f_a^4}  \\
\frac{d Y_{\chi}}{d T}
&=& -\frac{16\times 45}{2 \times  \pi^7 \times 1.66 }\dfrac{C_{G}^2 C_{\chi}^2 M_{\chi}^2 M_{pl}}{f_a^4\, g_{*}^s \, \sqrt{g_{\rho}}}
\label{eq:app_int}
\end{eqnarray}
Here ,$Y_{\chi}=n_{\chi}/s$ and $s$ is the entropy density.
Integrating eq.\eqref{eq:app_int} from $T=T_{RH}$ to $T=0$ we will get,
\begin{equation}
    Y_{\chi} \sim \dfrac{C_{G}^2 C_{\chi}^2 M_{\chi}^2 M_{pl} }{ f_a^4} T_{RH}
    \label{eq:cfg}
\end{equation}

\subsection{Non-thermalization of $gg\to \chi \chi$}
\label{sub:th_g}
The interaction rate is given by 
\begin{equation}
 \Gamma_{gg \to \chi \chi}= \frac{1}{n_g^{eq}}~ C[\mathcal{F}_g]   
\end{equation}
\begin{equation}
 n_g^{eq}(T)=\frac{3 \zeta(3)}{\pi^2 } T^3,
\end{equation}
where,
$\zeta(3)$ is the Riemann zeta function.
From eq.\eqref{eq:cf3} we get,
\begin{equation}
 \Gamma_{gg \to \chi \chi}=  \frac{ T^3}{0.36 \pi^5}\dfrac{16 C_{G}^2 C_{\chi}^2 M_{\chi}^2}{f_a^4} 
 \label{eq:intg}
\end{equation}
\begin{equation}
   H(T)= \frac{1}{M_{pl}} \sqrt{\frac{8\pi \rho}{3}} 
   \label{eq:hubble}
\end{equation}
where, $M_{pl}=1.22\times 10^{19}$ GeV is the Planck mass and $\rho$ is radiation energy density given by, 
\begin{equation}
   \rho=\frac{\pi^2}{30} g_{\rho} T^4 
\end{equation}
From eq.\eqref{eq:therm} comparing $\Gamma_{gg \to \chi \chi}(T_{RH})$ and $H(T_{RH})$ we get, 
\begin{equation}
    T_{RH}\lesssim  \left(\frac{C_G}{f_a} \right)^{-2} \left(\frac{C_{\chi}}{f_a} \right)^{-2} M_{\chi}^{-2} M_{pl}^{-1}.
\end{equation}
\section{UV freeze-in from fermion interactions}
\label{apxcd}
The amplitude for the process $\Bar{f}+f \rightarrow \Bar{\chi} +\chi+h$
is given by,
\begin{equation}
 |M|^2_{\rm \Bar{f}f-\Bar{\chi}\chi h}= 8 ~y_f^2 ~N_f^c ~\dfrac{C_{f}^2 C_{\chi}^2 M_{\chi}^2 s}{f_a^4 (s-M_a^2)^2} (4 s-8 m_f^2) ,    
\end{equation}
where, $p_{1,2}$ are the momentum of incoming fermions and $p_{3,4}$ are the momentum of outgoing particles.
$N_f^c$ is the color factor associated with quarks.
Consider the process: $\Bar{f}+f \rightarrow \Bar{\chi} +\chi+h$
The Boltzmann equation is given by,
\begin{equation}
  \dot{n}_{\chi} + 3 H n_{\chi} = \int d\Pi_f d\Pi_f d\Pi_h d\Pi_{\chi} d\Pi_{\chi} \delta^4(p_1+ p_2 -p_3-p_4-p_5) |M|^2_{\rm \Bar{f}f \to \chi\chi h} \mathcal{F}_f(p_1) \mathcal{F}_f(p_2) 
  \label{eq:cf_f}
\end{equation}
Here again we assume initial bath particles follow MB distribution, $\mathcal{F}_f \sim e^{-E_f/T}$. 
The term in the R.H.S of eq \eqref{eq:cf} is the collision term $C[\mathcal{F}_f]$.
We can simplify $C[\mathcal{F}_f]$ starting from the initial state phase space
\begin{equation}
d^3p_1 d^3 p_2 = (4 \pi |p_1| E_1 dE_1) (4 \pi |p_2| E_2 dE_2) \left(\frac{1}{2} \cos{\theta}\right)   
\end{equation}
Following the notation of ref.\cite{Gondolo:1990dk,Elahi:2014fsa} we redefine following parameters as:
\begin{eqnarray}
 E_+=E_1+E_2,~~
 E_-=E_1-E_2,~~
 s=2 M_f^2 +2 E_1 E_2 -2 |p_1||p_2|\cos{\theta}.
\end{eqnarray}
and after a bit algebra the collision term $C[\mathcal{F}_f]$ further reduces to
\begin{eqnarray}
\nonumber
    C[\mathcal{F}_f] = \int ds  && \frac{T}{(2 \pi)^4} \frac{\sqrt{s}}{4}~K_1\left(\frac{\sqrt{s}}{T} \right)  ~d\Pi_h d\Pi_{\chi} d\Pi_{\chi} \times\\&& \times \delta^4(p_1+ p_2 -p_3-p_4-p_5) ~|M|^2_{\rm \Bar{f}f \to \chi\chi h}.
    \label{eq:cfi}
\end{eqnarray}

Now, more challenging task remains in solving the final state phase space integral. In c.o.m frame $\Vec{p_1}+\Vec{p_2}=0$.  
Using the delta function we can write $\Vec{p_3}+\Vec{p_4}+\Vec{p_5}=0$
and following ref.\cite{Elahi:2014fsa} we transform the collision term as
\begin{equation}
 C[\mathcal{F}_f]= \frac{T}{16~(2\pi)^7} \int_{0}^1 dx_1 \int_{x_1}^1 dx_2 \int_{s_{min}}^{
 \infty} 
 ds ~ \frac{s^{3/2}}{4}~K_1\left(\frac{\sqrt{s}}{T} \right)    ~|M|^2_{\rm \Bar{f}f \to \chi\chi h}
\end{equation}
For process like $ff\to \chi \chi h$, $s_{min}=4 m_{\chi}^2$ (before EWSB).
After repeating the same steps as in eq.\eqref{eq:cf3}-eq.\eqref{eq:cfg} we get,
\begin{equation}
    Y_{\chi} \sim \dfrac{N_f^c~y_f~C_{f}^2 C_{\chi}^2 M_{\chi}^2 M_{pl} }{ f_a^4} T_{RH}
    \label{eq:cff}
\end{equation}

Following the same steps as in subsec. \ref{sub:th_g} we get the interaction rate similar like eq.\eqref{eq:intg} 
\begin{equation}
 \Gamma_{ff \to \chi \chi h}=  \frac{ T^3}{0.72 \pi^5}\dfrac{N_c C_{f}^2 C_{\chi}^2 M_{\chi}^2}{f_a^4} 
 \label{eq:intf}
\end{equation}

\bibliographystyle{JHEP}
\bibliography{ref1}

\end{document}